\documentclass[twocolumn,numberedappendix,apj]{openjournal} 

\usepackage{newtxtext,newtxmath}

\usepackage[T1]{fontenc}

\usepackage{graphicx}	
\usepackage{amsmath}     
\usepackage{amsfonts}    
\usepackage{amssymb}     
\usepackage{xcolor}      
\usepackage{hyperref}    
\hypersetup{colorlinks=true,urlcolor=[rgb]{0.5,0,0.6},filecolor=magenta,citecolor=[rgb]{0,0.2,0.7},linkcolor=magenta,}
\usepackage{orcidlink}   

\usepackage{lineno}     

\newcommand{\lcdm}{\ensuremath{\Lambda\textrm{CDM}}}
\newcommand{\wcdm}{\ensuremath{w\textrm{CDM}}}
\newcommand{\omegam}{\ensuremath{\Omega_{\mathrm{m}}}}

\newcommand{\omegab}{\ensuremath{\Omega_{\mathrm{b}}}}
\newcommand{\omegak}{\ensuremath{\Omega_{\mathrm{k}}}}
\newcommand{\Hnow}{\ensuremath{H_{0}}}
\newcommand{\hnow}{\ensuremath{h}}
\newcommand{\sigmaeight}{\ensuremath{\sigma_{8}}}
\newcommand{\seight}{\ensuremath{\hat{S}_{8}}}
\newcommand{\seightnohat}{\ensuremath{S_{8}}}
\newcommand{\ns}{\ensuremath{n_{\mathrm{s}}}}
\newcommand{\w}{\ensuremath{w}}
\newcommand{\Msun}{\ensuremath{\mathrm{M}_{\odot}}}
\newcommand{\Msunh}{\ensuremath{h^{-1}\mathrm{M}_{\odot}}}
\newcommand{\Mpc}{\ensuremath{\mathrm{Mpc}}}
\newcommand{\Mpch}{\ensuremath{h^{-1}\mathrm{Mpc}}}

\newcommand{\Mfiveoo}{\ensuremath{M_{500\mathrm{c}}}}

\newcommand{\Rtwooo}{\ensuremath{R_{200\mathrm{c}}}}
\newcommand{\Mtwooo}{\ensuremath{M_{200\mathrm{c}}}}
\newcommand{\ctwooo}{\ensuremath{c_{200\mathrm{c}}}}
\newcommand{\redshift}{\ensuremath{z}}
\newcommand{\mass}{\ensuremath{M}}
\newcommand{\dif}{\ensuremath{\mathrm{d}}}

\newcommand{\Nclusters}{\ensuremath{N_{\mathrm{tot}}}}

\newcommand{\planck}{\emph{Planck}}

\newcommand{\erosita}{\emph{eROSITA}}

\newcommand{\rich}{\ensuremath{N_{\mathrm{mem}}}}

\newcommand{\zcl}{\ensuremath{z_{\mathrm{cl}}}}
\newcommand{\zs}{\ensuremath{z_{\mathrm{s}}}}
\newcommand{\rshear}{\ensuremath{\gamma_{+}}}

\newcommand{\convergence}{\ensuremath{\kappa}}

\newcommand{\lensingw}{\ensuremath{w}}

\newcommand{\Sigmam}{\ensuremath{{\Sigma}_{\mathrm{m}}}}

\newcommand{\bwl}{\ensuremath{b_{\mathrm{WL}}}}
\newcommand{\mwl}{\ensuremath{M_{\mathrm{WL}}}}
\newcommand{\sigmacrit}{\ensuremath{\Sigma_{\mathrm{crit}}}}
\newcommand{\lensingeff}{\ensuremath{\beta}}
\newcommand{\rs}{\ensuremath{r_{\mathrm{s}}}}
\newcommand{\rhos}{\ensuremath{\rho_{\mathrm{s}}}}
\newcommand{\snr}{\ensuremath{\nu}}
\newcommand{\numin}{\ensuremath{\nu_{\mathrm{min}}}}
\newcommand{\mkappanohat}{\ensuremath{M_{\kappa}}}
\newcommand{\mkappa}{\ensuremath{\hat{M}_{\kappa}}}
\newcommand{\thetas}{\ensuremath{\theta_{\mathrm{s}}}}
\newcommand{\sigmashape}{\ensuremath{\sigma_{\kappa}}}
\newcommand{\sigmaconcen}{\ensuremath{\sigma_{c}}}
\newcommand{\Ufilter}{\ensuremath{U}}
\newcommand{\Qfilter}{\ensuremath{Q}}
\newcommand{\rdd}{\ensuremath{R}}
\newcommand{\rddd}{\ensuremath{r}}
\newcommand{\comp}{\ensuremath{\mathcal{C}}}
\newcommand{\deltasel}{\ensuremath{\Delta_{\mathrm{s}}}}

\newcommand{\Deltaz}{\ensuremath{\Delta z}}
\newcommand{\mpiv}{\ensuremath{M_{\mathrm{piv}}}}
\newcommand{\zpiv}{\ensuremath{z_{\mathrm{piv}}}}

\newcommand{\Awl}{\ensuremath{A_{\mathrm{WL}}}}
\newcommand{\Bwl}{\ensuremath{B_{\mathrm{WL}}}}
\newcommand{\deltawl}{\ensuremath{\delta_{\mathrm{WL}}}}
\newcommand{\gammawl}{\ensuremath{\gamma_{\mathrm{WL}}}}
\newcommand{\sigmawl}{\ensuremath{\sigma_{\mathrm{WL}}}}

\newcommand{\Brich}{\ensuremath{B_{N_{\mathrm{mem}}}}}

\newcommand{\sigmarich}{\ensuremath{\sigma_{N_{\mathrm{mem}}}}}
\newcommand{\omegamDefault}{\ensuremath{ 0.50^{+0.28}_{-0.24} }}
\newcommand{\sigmaeightDefault}{\ensuremath{ 0.685^{+0.161}_{-0.088} }}
\newcommand{\seightDefault}{\ensuremath{ 0.835^{+0.041}_{-0.044} }}
\newcommand{\seightNormDefault}{\ensuremath{ 0.993^{+0.084}_{-0.126} }}

\newcommand{\seightClusterz}{\ensuremath{ 0.849^{+0.043}_{-0.049} }}

\newcommand{\seightMZdep}{\ensuremath{ 0.839^{+0.043}_{-0.049} }}

\newcommand{\seightLowNumin}{\ensuremath{ 0.852^{+0.039}_{-0.040} }}

\newcommand{\seightLowSN}{\ensuremath{ 0.814^{+0.102}_{-0.090} }}

\newcommand{\seightHghSN}{\ensuremath{ 0.820\pm 0.047 }}

\newcommand{\seightWcdm}{\ensuremath{ 0.836^{+0.043}_{-0.044} }}

\newcommand{\percent}{\ensuremath{\%}}
\newcommand{\appropto}{\mathrel{\vcenter{
                        \offinterlineskip\halign{\hfil$##$\cr
                        \propto\cr\noalign{\kern2pt}\sim\cr\noalign{\kern-2pt}}}}}

\newcommand{\vect}[1]{\boldsymbol{\mathbf{#1}}}
\DeclareMathOperator\arctanh{arctanh}

\begin{document}


\title{
Weak-lensing Shear-selected Galaxy Clusters from the Hyper Suprime-Cam Subaru Strategic Program: \\
II. Cosmological Constraints from the Cluster Abundance
\vspace{-4.5em}
}
\shorttitle{Shear-selected cluster cosmology}
\shortauthors{Chiu et al.}

\author{I-Non~Chiu$^{1}$\thanks{E-mail:inchiu@phys.ncku.edu.tw}\orcidlink{0000-0002-5819-6566}}
\author{Kai-Feng~Chen$^{2,3}$\orcidlink{0000-0002-3839-0230}}
\author{Masamune~Oguri$^{4,5}$\orcidlink{0000-0003-3484-399X}}
\author{Markus~M.~Rau$^{6,7}$\orcidlink{0000-0003-3709-1324}}
\author{Takashi~Hamana$^{8}$}
\author{Yen-Ting~Lin$^{9}$\orcidlink{0000-0001-7146-4687}}
\author{Hironao~Miyatake$^{10,11,12}$\orcidlink{0000-0001-7964-9766}}
\author{Satoshi~Miyazaki$^{13}$\orcidlink{0000-0002-1962-904X}}
\author{Surhud~More$^{14,12}$\orcidlink{0000-0002-2986-2371}}
\author{Tomomi~Sunayama$^{15,12}$\orcidlink{0009-0004-6387-5784}}
\author{Sunao~Sugiyama$^{16,12}$\orcidlink{0000-0003-1153-6735}}
\author{Masahiro~Takada$^{12}$\orcidlink{0000-0002-5578-6472}}
%

\affiliation{$^{1}$Department of Physics, National Cheng Kung University, No.1, University Road, Tainan City 70101, Taiwan}
\affiliation{$^{2}$MIT Kavli Institute, Massachusetts Institute of Technology, Cambridge, MA 02139, USA}
\affiliation{$^{3}$Department of Physics, Massachusetts Institute of Technology, Cambridge, MA 02139, USA}
\affiliation{$^{4}$Center for Frontier Science, Chiba University, 1-33 Yayoi-cho, Inage-ku, Chiba 263-8522, Japan}
\affiliation{$^{5}$Department of Physics, Graduate School of Science, Chiba University, 1-33 Yayoi-Cho, Inage-Ku, Chiba 263-8522, Japan}
\affiliation{$^{6}$High Energy Physics Division, Argonne National Laboratory, Lemont, IL 60439, USA}
\affiliation{$^{7}$McWilliams Center for Cosmology, Department of Physics, Carnegie Mellon University, 5000 Forbes Avenue, Pittsburgh, PA 15213, USA}
\affiliation{$^{8}$National Astronomical Observatory of Japan, National Institutes of Natural Sciences, Mitaka, Tokyo 181-8588, Japan}
\affiliation{$^{9}$Academic Sinica Institute of Astronomy and Astrophysics, 11F of AS/NTU Astronomy-Mathematics Building, No.1, Sec. 4, Roosevelt Rd, Taipei 106216, Taiwan}
\affiliation{$^{10}$Kobayashi-Maskawa Institute for the Origin of Particles and the Universe (KMI), Nagoya University, Nagoya, 464-8602, Japan}
\affiliation{$^{11}$Institute for Advanced Research, Nagoya University, Nagoya 464-8601, Japan}
\affiliation{$^{12}$Kavli Institute for the Physics and Mathematics of the Universe (WPI), The University of Tokyo Institutes for Advanced Study (UTIAS), The University of Tokyo, Chiba 277-8583, Japan}
\affiliation{$^{13}$Subaru Telescope, National Astronomical Observatory of Japan, 650 North Aohoku Place Hilo, HI 96720, USA}
\affiliation{$^{14}$The Inter-University Centre for Astronomy and Astrophysics, Post bag 4, Ganeshkhind, Pune 411007, India}
\affiliation{$^{15}$Department of Astronomy and Steward Observatory, University of Arizona, 933 North Cherry Avenue, Tucson, AZ 85719, USA}
\affiliation{$^{16}$Center for Particle Cosmology, Department of Physics and Astronomy, University of Pennsylvania, Philadelphia, PA 19104, USA}

%
%

\begin{abstract}
We present cosmological constraints using the abundance of weak-lensing shear-selected galaxy clusters in the Hyper Suprime-Cam (HSC) Subaru Strategic Program.
The clusters are selected on the aperture-mass maps constructed using the three-year (Y3) weak-lensing data with an area of $\approx500~$deg$^2$, resulting in a sample size of $129$ clusters with high signal-to-noise ratios 
$\snr\geq4.7$.
Owing to the deep, wide-field, and uniform imaging of the HSC survey, this is by far the largest sample of shear-selected clusters, for which the selection solely depends on gravity and is free from any assumptions about the dynamical state and complex baryon physics. 
Informed by the optical counterparts, the shear-selected clusters span a redshift range of $\redshift\lesssim0.7$ with a median of $\redshift\approx0.3$.
The lensing sources are securely selected at $\redshift\gtrsim0.7$ with a median of $\redshift\approx1.3$, leading to nearly zero cluster member contamination.
We carefully account for 
(1) the bias in the photometric redshift of sources, 
(2) the bias and scatter in the weak-lensing mass using a simulation-based calibration, 
and (3) the measurement uncertainty that is directly estimated on the aperture-mass maps using an injection-based method developed in a companion paper \citep{chen24}.
In a blind analysis, the fully marginalized posteriors of the cosmological parameters are obtained as 
$\omegam = \omegamDefault$, $\sigmaeight = \sigmaeightDefault$, $\seight\equiv\sigmaeight\left(\omegam/0.3\right)^{0.25} = \seightDefault$, and 
$\sigmaeight\sqrt{\omegam/0.3} = \seightNormDefault$ in a flat \lcdm\ model.
Our results are robust against the systematic uncertainties of the weak-lensing mass bias, the photo-\redshift\ bias, the fitting formula of the halo mass function, the different selection criteria ($\snr\geq4.3$, $\snr\geq5.3$, and $4.7\leq\snr<5.3$), and the individual subfields.
We compare our cosmological constraints with other studies, including those based on cluster abundances, galaxy-galaxy lensing and clustering, and Cosmic Microwave Background observed by \planck, and find good agreement at levels of $\lesssim2\sigma$.
This work realizes a cosmological probe utilizing weak-lensing shear-selected clusters and paves the way forward in the upcoming era of wide-field sky surveys.
\end{abstract}


\maketitle


%
%

\section{Introduction}
\label{sec:intro}

Galaxy clusters are powerful in probing cosmology, because they are peaks of the cosmic density field and their populations over time closely trace the structure formation of the Universe.
In particular, the number density (i.e., the abundance) of galaxy clusters as a function of mass is sensitive to the mean matter density \omegam\ and the r.m.s. fluctuation \sigmaeight\ of the matter-density field at a scale of $8\Mpch$ \citep[see reviews in][]{haiman01,allen11}.
Furthermore, measurements of cluster abundances at different redshifts directly constrain the history of the structure growth and, hence, the nature of dark energy \citep{weinberg13,huterer15}.

Two key challenges to executing a cosmological study using the abundance of galaxy clusters (hereafter clusters) are (1) the construction of a sizable sample with a well-understood selection function, and (2) the accurate determination of the cluster mass.
For the former, owing to the deployments of large-sky surveys, large samples of clusters over a wide range of mass and redshift have been constructed in multiple wavelengths, including those 
detected in millimeter surveys \citep{bleem15,bleem20,hilton21} via the Sunyaev-Zel’dovich effect \citep[SZE;][]{sunyaev72}, 
identified in optical imaging by the overdensity of galaxies \citep{gladders07,rykoff14,oguri14,bellagamba18}, 
selected by the emission of the intracluster medium (ICM) in X-rays \citep{bohringer04,vikhlinin09a,adami18,klein19,liu22}, 
and constructed in a combination of the above \citep{klein19,klein22,klein23,hernandez-lang23}.
Meanwhile, the latter task has been realized by the technique of weak gravitational lensing (hereafter weak lensing or WL), enabling a direct probe to the total mass of a cosmic object by leveraging only the theory of General Relativity.
Over the last two decades, there has been tremendous progress and success in the weak-lensing mass calibration of clusters \citep{clowe00,mandelbaum06,okabe10,umetsu14,vonderlinden14a,gruen14,hoekstra15,schrabback18,schrabback21}, especially those utilizing wide-field imaging surveys \citep{simet17,mcclintock19,murata19,bellagamba19,umetsu20,chiu22}.
Nowadays, it has been not only a norm but also a necessity to deliver robust cosmological constraints from clusters by combining the abundance and the weak-lensing mass calibration \citep{oguri11b,mantz15,bocquet19,costanzi21,chiu23,sunayama24,bocquet24,fumagalli24,ghirardini24}.

A well-received strategy in cluster cosmology is to construct a large cluster sample selected by a baryon-based observable (e.g., the X-ray emission from ICM, or the stellar light of member galaxies) that is then associated with the cluster mass through the so-called ``observable-to-mass relation (OMR)''.
By the weak-lensing mass calibration of the OMR, as a ``post-selection'' analysis, the constraint on cosmology is obtained 
by comparing the prediction from the halo mass function and the observed abundance in the observable space.
Inevitably, this relies on assumptions about the baryonic properties of clusters, which are extremely difficult to predict.
For example, a gravity-only system predicts a self-similar OMR that the thermal bremsstrahlung X-ray luminosity of the ICM scales with the cluster mass to a power of $4/3$.
However, the observed X-ray luminosity-to-mass relation reveals a significantly steeper mass trend than the self-similar slope \citep[][and references therein]{bulbul19,chiu22}.
This strongly suggests that the presence of non-gravitational processes (e.g., the feedback from active galactic nuclei) plays an important role in the baryonic observable of clusters.
Moreover, the diversity in the cluster formation history results in the intrinsic scatter of an observable with an unknown amplitude at a fixed cluster mass.
These complexities in a baryon-based observable result in a selection function that must be modelled empirically \citep{grandis20,chiu23} or calibrated against extensive simulations \citep[c.f.,][]{liuteng22,seppi22}.

Wide-field and deep weak-lensing surveys provide an alternative and promising way to cluster cosmology.
The weak-lensing data with a high source number density significantly increase the resolution of the resulting aperture-mass maps, in which clusters can be clearly identified as lensing peaks \citep{schneider96,jain00,white02a,hamana04,hennawi05}.
In this way, one can select clusters purely based on gravity and independently of any baryonic assumptions.
This results in a ``gravity-selected'' (or ``shear-selected'') sample with a selection that is directly related to the cluster mass.
In fact, the selection observable of shear-selected clusters is effectively the weak-lensing mass, such that the modelling of the abundance and the mass calibration is performed based on a single observable---the weak-lensing signal-to-noise ratio \snr---without relying on an additional baryon-based mass proxy.
This not only largely reduces the number of nuisance parameters in the modelling, but also gives a ``self-contained'' and ``baryonic-tracer-free'' probe to cosmology,
as the most attractive and unique advantage over other cluster samples \citep{wittman01,miyazaki02,miyazaki07,schirmer07,gavazzi07,hamana15}.

The major task to constrain cosmology using shear-selected clusters is to establish the mapping between the cluster halo mass \mass, which we use to parameterize the halo mass function, and the observed lensing signal-to-noise ratio \snr, which is the selection observable and subject to the measurement uncertainty of the weak-lensing mass \mwl.
The mapping is twofold, as follows. 
The first is the mapping between the halo mass \mass\ and the weak-lensing mass \mwl, which accounts for the intrinsic diversity of the halo mass profile at a fixed \mass. 
By leveraging the facts that clusters are dominated by dark matter and that their average mass profiles appear highly self-similar \citep[see e.g.,][]{wang20}, the scatter and bias of the weak-lensing mass can be calibrated against numerical simulations in a straightforward and empirical way \citep[e.g.,][]{grandis21}.
Note that the feature of ``being calibratable'' using simulations makes the shear-selected cluster cosmology extremely comprehensible compared to others using a baryon-based tracer.
The second mapping is between the weak-lensing mass \mwl\ and the observable \snr, which is completely attributed to the measurement uncertainty.
In a companion paper \citep{chen24}, we developed a novel method to quantify the measurement uncertainty 
directly on the observed aperture-mass maps.

In this work, we carry out a cosmological study using the abundance of shear-selected clusters.
The shear-selected sample is constructed using the latest weak-lensing year-three (Y3) data from the Hyper Suprime-Cam Subaru Strategy Program \citep[the HSC survey;][]{aihara18a}.
We stress that the HSC survey provides the weak-lensing data with the highest source density at the achieved area to date ($\approx21$ galaxies/arcmin$^2$ over the HSC-Y3 footprint with a total area of $\approx500$~deg$^2$), as the only existing weak-lensing survey with sufficiently deep imaging and large area to carry out such a cosmological analysis.
The first sample of shear-selected clusters over a hundred-square-degree footprint area was built
in the HSC first-year data set \citep{miyazaki18b}, followed by the HSC-Y3 sample presented in \cite{oguri21} with more advanced map filtering and source selections.
This study is in distinction to weak-lensing peak statistics \citep[e.g.,][]{fan10,dietrich10,berge10,liux15,liu15,kacprzak16,martinet18,shan18,liu19,liu23,marques24} 
in two perspectives, namely
(1) the aperture-mass maps are smoothed by a kernel specifically optimized for clusters with the goal of minimizing the noise from large-scale structures and other associating systematics, and that
(2) the lensing sources are strictly selected to avoid cluster member contaminations.
In-depth examinations of the adopted smoothing kernel and the source selection are given in \cite{oguri21}, which we refer readers to for more details.

The paper is organized as follows.
We introduce the data in Section~\ref{sec:data}.
The methodology is described in detail in Section~\ref{sec:methodology}, while the modelling is presented in Section~\ref{sec:modelling}.
We show
the results in Section~\ref{sec:results} and make conclusions in Section~\ref{sec:conclusions}.
Throughout this paper, the uncertainties are quoted as the $68\percent$ confidence levels ($1\sigma$).
Unless stated otherwise, we made the following conventions that (1) the definition of the cluster mass is defined as the mass enclosed by a sphere wherein the average mass density is
$200$ times the critical density at the cluster redshift,
and that (2) the halo mass is symbolized by \mass\ and interchangeable with \Mtwooo.
The notation $\mathcal{N}\left(x,y^2\right)$ ($\mathcal{U}\left(x,y\right)$) stands for a Gaussian distribution with a mean $x$ and a standard deviation $y$ (a flat interval between $x$ and $y$).

%
%

\section{Data}
\label{sec:data}

Brief descriptions of the HSC data sets
are provided in this section.
We note that the same data sets have been used to deliver robust constraints on cosmology based on cosmic shears \citep{li23,dalal23,more23,miyatake23,sugiyama23}.

\subsection{Weak-lensing shape catalogs}
\label{eq:wldata}

We use the shape catalogs from the HSC-Y3 weak-lensing data sets that are fully described in \cite{li22}.
In short, the $i$-band imaging observed by the HSC survey from 2014 to 2019 was collected and used to produce the shape catalogs with a median seeing of $\approx0.59$~arcsec and a magnitude cut at $i\leq24.5$~mag, resulting in an effective area of $\approx500$~deg$^2$ consisting of six subfields (GAMA09H, GAMA15H, XMM, VVDS, HECTOMAP, WIDE12H).
The effective number density of galaxies in the shape catalog reaches $\approx20$~arcmin$^{-2}$.
The shape measurement is rigorously calibrated against image simulations following the prescription in \cite{mandelbaum18a}, delivering a multiplicative bias $\delta m$ at a level of $|\delta m| < 9\times10^{-3}$.
Detailed examinations of the shape catalogs in \cite{li22} demonstrate that various null tests are consistent with zero, providing the shape measurements that are sufficiently accurate for this work.

\subsection{Photometric redshifts}
\label{sec:photoz}

The photometric redshifts (hereafter photo-\redshift) of individual galaxies observed in the HSC-Y3 data sets are estimated using the $grizY$ broadband photometry.
The methodology to derive the photo-\redshift\ is described in depth in \cite{tanaka18}.
For this work, we use the photo-\redshift\ from the HSC Public Data Release 2 \citep{nishizawa20}.
Specifically, we use the photo-\redshift\ estimated by the ``Direct Empirical Photometric method'' code \citep[\texttt{DEmP};][]{hsieh14}, which utilizes a machine-learning-based algorithm.
In \cite{nishizawa20}, the HSC photo-\redshift\ measurements obtained by \texttt{DEmP} has been fully examined using spectroscopic redshifts, delivering a performance of the bias, scatter, and the outlier fraction at levels of
$|\Delta\redshift| \approx 0.003\left(1 + \redshift\right)$,
$\sigma_{\Delta\redshift} \approx 0.019\left(1 + \redshift\right)$, and 
$\approx5.4\percent$, respectively, for galaxies with magnitude $i\leq24.5$~mag.
It is worth mentioning that the photo-\redshift\ estimated by \texttt{DEmP} has been widely used in not only the studies of galaxy clusters \citep{chiu20,chiu22,chiu23} but also the HSC first-year analyses of cosmic shears \citep{hikage19,hamana20}, demonstrating the reliability of the photo-\redshift\ measurements for the purposes of this work.

%
%

\section{Methodology}
\label{sec:methodology}

In this section, we provide a review of weak lensing in the perspective of shear-selected clusters (section~\ref{sec:wlbasics}) and 
the map filtering (section~\ref{sec:filters}), the selection of the source sample (section~\ref{sec:source_selection}), the practical procedures in producing the aperture-mass maps (section~\ref{sec:practical}), and the construction of the shear-selected cluster sample (section~\ref{sec:clustersample}).

\subsection{Weak-Lensing basics}
\label{sec:wlbasics}

Lights received by observers are deflected due to the presence of cosmic structures along the line of sight, as the effect of gravitational lensing \citep[see e.g.,][for a complete review]{bartelmann01,umetsu20b}.
The strength of gravitational lensing is determined by the distances between the lens-observer, source-observer, and lens-source pairs.
Considering a galaxy cluster as a single-thin lens at redshift \zcl, the lensing strength of a source at redshift \zs\ is described by the so-called ``critical surface mass density'',
\begin{equation}
\label{eq:sigmacrit}
\sigmacrit\left(\zcl,\zs\right) = \frac{c^2}{4\pi G} \frac{1}{ D_{\mathrm{A}} \left(\zcl\right) \lensingeff\left(\zcl, \zs\right) } \, ,
\end{equation}
in which $c$ is the speed of light, $G$ is the Newton constant, and $\lensingeff\left(\zcl, \zs\right)$ is the lensing efficiency of the source at redshift \zs\ for the lens at redshift \zcl,
\begin{equation}
\lensingeff\left(\zcl, \zs\right) = \left\lbrace
\begin{array}{cc}
\frac{ D_{\mathrm{A}}\left(\zcl, \zs\right) }{ D_{\mathrm{A}}\left(\zcl\right) } &\mathrm{if~}\zs > \zcl \\ \nonumber
0 &\mathrm{if~}\zs \leq \zcl \nonumber
\end{array}
\right. \, ,
\end{equation}
where $D_{\mathrm{A}}\left(\redshift\right)$ and $D_{\mathrm{A}}\left({\redshift}_1, {\redshift}_2\right)$ are the angular diameter distances to the redshift \redshift\ and between the redshift pairs $\left( {\redshift}_1, {\redshift}_2\right)$, respectively. 

Given a source at redshift \zs, the convergence \convergence\ of the cluster at redshift \zcl\ and the sky location $\vect{\theta}$ is then expressed as 
\begin{equation}
\label{eq:kappa_def}
\convergence\left(\vect{\theta}\right) = \frac{ \Sigmam\left(\vect{\theta}\right)  }{ \sigmacrit\left(\zcl,\zs\right) } \, ,
\end{equation}
where $\Sigmam\left(\vect{\theta}\right)$ is the projected surface mass density of the cluster at the sky location $\vect{\theta}$, which is associated with the (projected) physical coordinate $\vect{\rdd}$ as $\vect{\rdd} = \vect{\theta} \times D_{\mathrm{A}}\left(\zcl\right)$.

In the approximation of a flat sky, the dimensionless aperture-mass map is obtained by convolving the convergence field with a kernel \Ufilter\ \citep{schneider96} as
\begin{equation}
\label{eq:mkappa_field}
\mkappanohat\left(\vect{\theta}\right) = 
\int \dif\vect{\theta}^{\prime}~
\convergence\left(\vect{\theta^{\prime}}\right) 
\Ufilter\left( | \vect{\theta^{\prime}} - \vect{\theta} | \right) \, .
\end{equation}
We require the kernel to be isotropic and ``compensated'' by satisfying 
\begin{equation}
\label{eq:compensated}
\int \dif \theta~\theta~\Ufilter\left( \theta \right) = 0 \, .
\end{equation}
In this way, because both shears \rshear\ and convergence \convergence\ are linear combinations of the second derivatives of the lensing potential, the convolution of the convergence field is written as a convolution of the shear field with a kernel \Qfilter, 
\begin{equation}
\label{eq:shear_field_convolution}
\mkappanohat\left(\vect{\theta}\right) = 
\int \dif\vect{\theta^{\prime}}
\rshear\left(\vect{\theta^{\prime}} ; \vect{\theta} \right)
\Qfilter\left( | \vect{\theta^{\prime}} - \vect{\theta} | \right) \, ,
\end{equation}
where $\rshear\left(\vect{\theta^{\prime}} ; \vect{\theta} \right)$ is the shear at $\vect{\theta^{\prime}}$ in the tangential direction defined with respect to $\vect{\theta}$, and the kernel \Qfilter\ is related to \Ufilter\ as
\begin{equation}
\label{eq:filter_conversion}
\Qfilter\left( \theta \right) = 
\frac{2}{\theta^2}
\int \dif\theta^{\prime} \theta^{\prime} U\left(\theta^{\prime}\right) - \Ufilter\left(\theta\right) \, .
\end{equation}

The signal-to-noise ratio \snr\ of the cluster at a sky location $\theta$ is then calculated as
\begin{equation}
\label{eq:snr_field}
\snr\left(\theta\right) = \frac{\mkappanohat\left(\theta\right)}{\sigmashape\left(\theta\right)} \, ,
\end{equation}
where \sigmashape\ is the shape noise of the aperture-mass map, as derived following the description in Section~\ref{sec:practical}.
The highest signal-to-noise ratio \snr\ is referred to as the lensing peak height of the cluster.
For a spherically symmetric cluster, the lensing peak occurs at the cluster center projected on the sky.

\subsection{Map filters}
\label{sec:filters}

In this work, as also fully explored in \cite{oguri21}, we use a truncated isothermal filter \citep{schneider96} that reads
\begin{equation}
\label{eq:ti20}
\Ufilter\left(\theta\right) = 
\left\lbrace
\begin{array}{cl}
1 
&~\mathrm{if}~\theta\leq\nu_1\theta_{\mathrm{R}} \\
\frac{1}{1 - \mathfrak{c} }  \left[ \frac{ \nu_1 \theta_{\mathrm{R}} }{ \sqrt{  \left(\theta - \nu_1 \theta_{\mathrm{R}}\right)^2 + \left(\nu_1 \theta_{\mathrm{R}}\right)^2 } } - \mathfrak{c} \right] 
&~\mathrm{if}~\nu_1\theta_{\mathrm{R}}\leq\theta\leq\nu_2\theta_{\mathrm{R}} \\
\frac{ \mathfrak{b} }{ {\theta_{\mathrm{R}}}^3 } \left( \theta_{\mathrm{R}} - \theta \right)^2 \left( \theta - \mathfrak{a} \theta_{\mathrm{R}}\right)
&~\mathrm{if}~\nu_2\theta_{\mathrm{R}}\leq\theta\leq\theta_{\mathrm{R}} \\
0
&~\mathrm{if}~\theta\geq\theta_{\mathrm{R}}
\end{array}
\right.
\, .
\end{equation}
The parameters $\mathfrak{a}$, $\mathfrak{b}$, and $\mathfrak{c}$ in equation~(\ref{eq:ti20}) needs to be solved for a given filter configuration characterized by the parameters of $\nu_1$, $\nu_2$, and $\theta_{\mathrm{R}}$.
This is done by requiring that the filter is compensated and smooth (i.e., \Ufilter\ and $\frac{\dif\Ufilter}{\dif\theta}$ are continuous) at $\theta = \nu_2\theta_{\mathrm{R}}$.
As identical to \cite{oguri21}, we choose $\nu_1 = 0.121$, $\nu_2 = 0.36$, and $\theta_{\mathrm{R}} = 16.6$~arcmin in this work\footnote{This filter was referred to as the \texttt{TI20} filter in \cite{oguri21}.}.

The truncated isothermal filter provides several major benefits.
First, it has the feature of $\Qfilter\left(\theta\right) = 0$ for $\theta \leq \nu_1\theta_{\mathrm{R}}$.
By carefully choosing $\nu_1$, we could remove the lensing signal of cluster inner regions that are severely dominated by various systematic effects, especially the contamination due to cluster member galaxies.
Second, the feature of $\Ufilter\left(\theta\right) < 0$ at $\nu_2\theta_{\mathrm{R}}\lesssim\theta\leq\theta_{\mathrm{R}}$ enables a local background subtraction in constructing the aperture-mass map.
Compared to a standard Gaussian filter, the local background subtraction significantly mitigates the lensing projection effects due to fluctuations of cosmic structures on a large scale.
Third, the filter is confined, i.e., $\Ufilter\left(\theta\right) = \Qfilter\left(\theta\right) = 0$ for $\theta \geq\theta_{\mathrm{R}}$.
This reduces the impacts from the outer regions that are not of interest, e.g., the boundary of survey footprints.

\subsection{Selections of lensing sources}
\label{sec:source_selection}

The sample of source galaxies used to construct the aperture-mass maps is selected based on their photometric redshifts \citep[the so-called P-cut selection; see also][]{oguri14,medezinski18a}, following the procedure in \cite{oguri21}.
Specifically, the individual redshift distributions $P\left(\redshift\right)$ estimated by the code \texttt{DEmP} \citep{hsieh14} are used, and a galaxy is selected as a source if its redshift distribution satisfies
\begin{equation}
\label{eq:source_selection}
\int\limits_{{\redshift}_{\mathrm{min}}}^{{\redshift}_{\mathrm{max}}} P\left(\redshift\right) \dif\redshift > P_{\mathrm{th}} \, ,
\end{equation}
in which we choose ${\redshift}_{\mathrm{min}} = 0.7$, ${\redshift}_{\mathrm{max}} = 7.0$, and $P_{\mathrm{th}} = 0.95$.
As a result, $\approx16$ millions out of $\approx36$ millions galaxies in the shape catalogs are selected as the lensing sources, corresponding to $\approx10$~galaxies per square arcmin.
The mean redshift of the source sample is $\approx1.28$.

The strict selection of the source sample is needed to remove the contamination in the lensing signals arising from non-background galaxies, especially cluster member galaxies.
The cluster member galaxies, if leaking into the source sample, decrease not only the lensing signal but also the shape noise \sigmashape.
The former reduction
is due to the dilution of the averaged distortion in the sample ellipticity, while the latter is a natural result that the variance ${\sigmashape}^2$ is inversely proportional to the source number density.
The combination of the two decrements leads to a net effect on the peak height that is challenging to be modelled.

In this work, we use a relatively conservative selection to choose
the source sample that are securely at redshift $\redshift\gtrsim0.7$, with the goal of completely avoiding the cluster member contamination.
As studied in \cite{oguri21} and \cite{chen24}, the cross match between the resulting shear-selected clusters and optically selected samples shows that nearly no shear-selected clusters are detected at redshift $\redshift\gtrsim0.7$.
This suggests that no cluster member contamination is expected, thus
enabling a contamination-free modelling in this work.

\subsection{Constructions of the maps in practice}
\label{sec:practical}

We construct the maps of the aperture mass, the shape noise and the signal-to-noise ratio from the HSC-Y3 weak-lensing shape catalogs by following the same procedure as in \cite{oguri21}, to which we refer readers for more details.
A brief summary is provided in the following. 

For the aperture-mass maps, we first create the shape maps in grids.
Specifically, after the source selection (see Section~\ref{sec:source_selection}) we employ a flat-sky approximation and project the footprint into a two-dimensional $x$--$y$ plane in a Cartesian coordinate by a tangent-plane projection\footnote{Note that we correct the artificial distortions in the resulting $x$--$y$ plane due to the tangent-plane projection.}.
Then, we grid the plane with a pixel scale of $\Delta\theta = 0.25$~arcmin.
In each grid, we derive the lensing-weight weighted ellipticity, response, multiplicative bias and additive bias.
Finally, we convolve the resulting shape map with the kernel \Qfilter\ to obtain the aperture-mass map by following the equation~(9) in \cite{oguri21}.
The convolution is carried out by using the fast Fourier transform.

The signal-to-noise ratio \snr\ of a lensing peak is obtained as the ratio of the convergence mass to the locally defined noise at the same sky position.
We construct the noise map, also referred to as the ``sigma map'', to acquire the locally defined noise.
To do so, we randomly rotate each source galaxy in the shape catalog and produce the corresponding randomized aperture-mass map following the same procedure described above.
This process is repeated for 500 times, resulting in 500 realizations of the randomized aperture-mass maps.
The sigma map is then derived as the standard deviation of these 500 randomized aperture-mass maps at each sky position.
Note that, by construction, the sigma map only records the local shape noise and does not include the information of cosmic shears arising from large-scale structures \citep[see Appendix in][]{oguri21}.
We account for the noise originated
from cosmic shears in the selection function \citep[see Section~\ref{sec:modelling_selection_function} and also][]{chen24}.

Finally, the signal-to-noise map is derived at each sky position as the ratio of the value in the aperture-mass map to that in the sigma map, i.e., equation~(\ref{eq:snr_field}).

\begin{figure}
\centering
\resizebox{0.47\textwidth}{!}{
\includegraphics[scale=1]{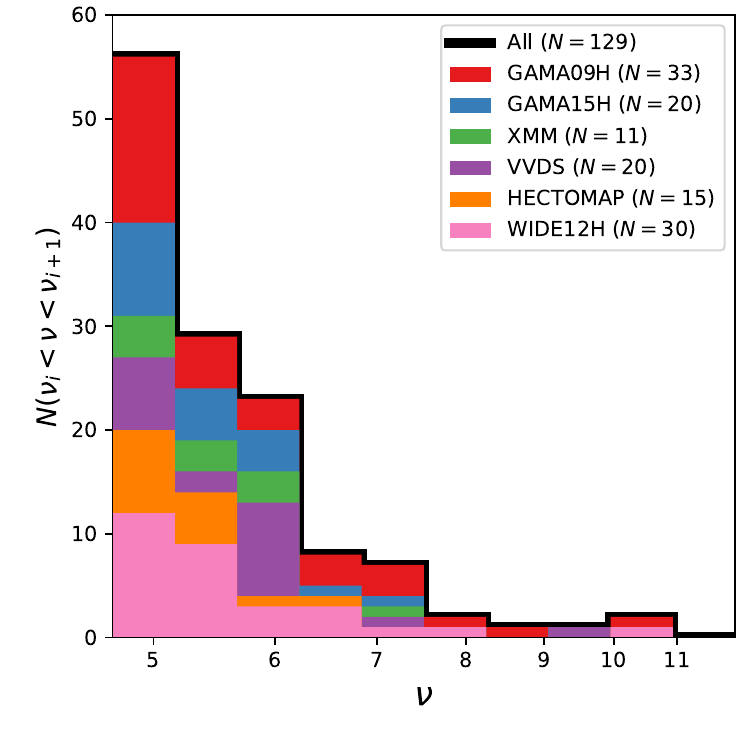}
}
\caption{
The observed numbers of shear-selected clusters.
The x-axis shows the signal-to-noise ratio \snr\ with the 10 logarithmic binning between $4.7$ and $12$.
The y-axis records the number of detected clusters in all footprints of the HSC-Y3 data (black) and subfields (the colors indicated in the upper-right corner).
}
\label{fig:sample}
\end{figure}

As identical to \cite{oguri21}, we mask the regions that have less than $0.5$ times the averaged source number density, which is derived from the number density map smoothed by a Gaussian kernel with a standard deviation of $8$~arcmin.
This is to remove the regions outside the survey footprint or in bright-star masks.
In addition, the pixels with the values in the sigma map that are $1.5$ times higher than the average value are also masked.
The averaged value of the shape noise \sigmashape\ is $\approx0.70$ among the six subfields with little variations (at a level of $\lesssim0.03$).

We note that these map productions are carried out separately for the six individual subfields.

\subsection{The sample of shear-selected clusters}
\label{sec:clustersample}

Shear-selected clusters are identified as the peaks in the signal-to-noise maps with the lensing peak height \snr, which is used as the mass proxy of individual clusters.
A pixel on the maps is identified as a weak-lensing peak if its signal-to-noise ratio is higher than all eight other adjacent pixels.
To construct a sample of shear-selected clusters, we simply impose a cut on \snr\ as
\begin{equation}
\label{eq:snr_cut}
\nu\geq\numin \, ,
\end{equation}
where \numin\ is the detection threshold.
In the default analysis, we use $\numin = 4.7$, resulting in $129$ shear-selected clusters over the HSC-Y3 footprint ($\approx500$~deg$^2$).
We therefore expect the Poisson error
of the sample at a level of $1/\sqrt{129}\approx9\percent$.
The resulting sample has the maximum signal-to-noise ratio of $\approx10.4$ with a median value of $\approx 5.3$.
Lowering (increasing) \numin\ to a value of $4.3$ ($5.3$) increases (decreases) the sample size to $207$ ($66$) clusters.
In a later analysis (see Section~\ref{sec:results}), we assess the systematic uncertainty of the final results raised from the different selections, as a test on the internal consistency. 

We show the resulting sample in Figure~\ref{fig:sample} with the individual subfields labelled by different colors.
The binning in \snr\ is $10$ equal-step logarithmic bins between $4.7$ and $12$.
The histograms of the cluster numbers are the data vector used to constrain cosmology.

%
%

\section{Modelling}
\label{sec:modelling}

We aim to constrain cosmological parameters based on the modelling of the observed number counts $N\left(\snr\right)$ as a function of the weak-lensing signal-to-noise ratio \snr.
We focus on the sample of peaks above a signal-to-noise threshold \numin, i.e., $\snr\geq\numin$.
The key to obtaining the cosmological constraints is to relate the observed $N\left(\snr\right)$ to the halo mass function predicted for a given cosmology.
In a forward-modelling approach, we first model the mass distribution of individual halos and interpret the resulting weak-lensing-inferred mass \mwl\ with a calibrated redshift distribution of the sources.
This establishes the association between the cluster mass and the observable \snr, enabling the constraints on cosmological parameters by comparing the model prediction and the data.
A novel perspective in this work is that we characterize the sample selection (which is $\snr\geq\numin$) together with the measurement uncertainty using semi-analytical injection simulations, leading to the selection function as a function of the halo profile quantities.
During the modelling, we account for the bias in the weak-lensing mass \mwl\ and the photometric redshift of the sources.

We perform the analysis blindly to avoid the confirmation bias.
Specifically,  we hide the absolute values of the cosmological constraints until the analysis passes all the systematic examinations.

In what follows, we describe the modelling of 
the cluster mass profile (Section~\ref{sec:modelling_massprofile}),
the source redshift distribution (Section~\ref{sec:modelling_Pz}),
the cluster abundance (Section~\ref{sec:modelling_abundance}),
the selection function (Section~\ref{sec:modelling_selection_function}), and
the systematic uncertainties, including the weak-lensing mass bias (Section~\ref{sec:modelling_bwl_bias}) and the source photo-\redshift\ bias (Section~\ref{sec:modelling_photoz_bias}).
The blinding strategy is given in Sections~\ref{sec:blinding}, and the statistical inference to the modelling of cosmology is described \ref{sec:modelling_statistics}.

\subsection{Cluster mass profiles}
\label{sec:modelling_massprofile}

In this work, we model the density profile of clusters using a spherical Navarro-Frenk-White \citep[hereafter NFW;][]{navarro97} model that reads
\begin{equation}
\label{eq:nfw}
\rho_{\mathrm{NFW}}\left(\rddd\right) = \frac{\rhos}{\left(\frac{\rddd}{\rs}\right)\left(1 + \frac{\rddd}{\rs}\right)^2} \, ,
\end{equation}
where $\rho_{\mathrm{NFW}}\left(\rddd\right)$ is the mass density at the cluster radius \rddd, \rhos\ is the normalization, and \rs\ is the scale radius.
With the NFW model, the projected surface mass density profile $\Sigmam\left(\rdd\right)$ at a projected radius \rdd\ can be evaluated analytically as
\begin{multline}
\label{eq:kappa_profile}
\Sigmam\left(\rdd\right) \equiv 
\int\limits_{-\infty}^{\infty} \rho_{\mathrm{NFW}}\left(\sqrt{ {\rdd}^2 + \mathfrak{z}^2 } \right)~\dif\mathfrak{z} \\
= 2\rhos\rs \times 
\left\lbrace
\begin{array}{ll}
\frac{1}{1 - x^2} \left[-1 + \frac{2}{\sqrt{1-x^2}} \arctanh\left( \sqrt{\frac{1 - x}{1 + x}} \right) \right]   &\mathrm{if}~x<1 \,  \\
\frac{1}{3}  &\mathrm{if}~x=1 \, \\
\frac{1}{x^2 - 1} \left[1  - \frac{2}{\sqrt{x^2-1}} \arctan \left(\sqrt{\frac{x-1}{x+1}} \right) \right]   &\mathrm{if}~x>1 \, 
\end{array}
\right.
\, ,
\end{multline}
where $x\equiv \frac{R}{\rs}$.
Given a cluster mass \mass\ and a halo concentration \ctwooo, which is defined as the ratio of the cluster radius \Rtwooo\ to the scale radius \rs, 
\[
\ctwooo \equiv \frac{\Rtwooo}{\rs} \, ,
\]
one uniquely determines the NFW model and the projected surface mass density profile $\Sigmam\left(\rdd\right)$.

In this work, we use the location of the observed lensing peak as the cluster center.
Given the inner smoothing scale of the filter ($\theta\lesssim2~\mathrm{arcmin}$), we do not expect significant impact from the miscentering on the aperture mass.
The effect of miscentering is accounted for in determining the measurement uncertainty in the observed lensing peak height \snr\ through the detection of synthetic clusters injected into the observed aperture-mass maps (see Section~\ref{sec:modelling_selection_function}).
In fact, we find that majority ($\gtrsim80\percent$) of injected clusters have the central offset to the true center at a scale $\lesssim2~\mathrm{arcmin}$ \citep[see also][]{chen24}.
This suggests that the miscentering effect is subdominant.
On the other hand, we also examine another model of the halo density profile \citep{oguri11}, which accounts for the two-halo term analytically, and find negligible difference.
We therefore conclude that the default NFW model provides a sufficiently accurate description for the purpose of this work.

\subsection{Source redshift distributions}
\label{sec:modelling_Pz}

To compute the dimensionless convergence \convergence\ from the projected surface mass profile \Sigmam\ of a cluster, the redshift of sources must be utilized to calculate the critical surface mass density \sigmacrit.

In this work, the redshift distribution estimated by the code \texttt{DEmP} is used to infer the redshift of the source galaxies.
Specifically, the average of the inverse critical surface mass density for a given cluster redshift \zcl\ is determined as
\begin{equation}
\label{eq:calc_invsigmacrit}
\left\langle\frac{1}{\sigmacrit}\right\rangle = 
\int
\frac{1}{ \sigmacrit\left(\zcl,\redshift\right) }~
P_{\mathrm{s}}\left(\redshift\right) \dif\redshift \, ,
\end{equation}
where $P_{\mathrm{s}}\left(\redshift\right)$ is the stacked and lensing-weight-weighted redshift distribution (hereafter the stacked redshift distribution) and is obtained as
\begin{equation}
\label{eq:wPz}
P_{\mathrm{s}}\left(\redshift\right) = \frac{ 
\Sigma_{i} {\lensingw}_{i} P_{i}\left(\redshift\right)
}{
\Sigma_{i} {\lensingw}_{i}
}
\, ,
\end{equation}
in which $P_{i}\left(\redshift\right)$ and ${\lensingw}_{i}$ are the redshift distribution and the lensing weight of the $i$-th source galaxy, respectively.
In this way, given an NFW model at a cluster redshift, we can estimate the convergence field as
\begin{equation}
\label{eq:estimate_kappa}
\convergence = \left\langle\frac{1}{\sigmacrit}\right\rangle \times \Sigmam
\end{equation}
and, hence, predict the dimensionless aperture mass peak \mkappa\ at the position of the peak using equation~(\ref{eq:mkappa_field}).

\begin{figure}
\centering
\resizebox{0.48\textwidth}{!}{
\includegraphics[scale=1]{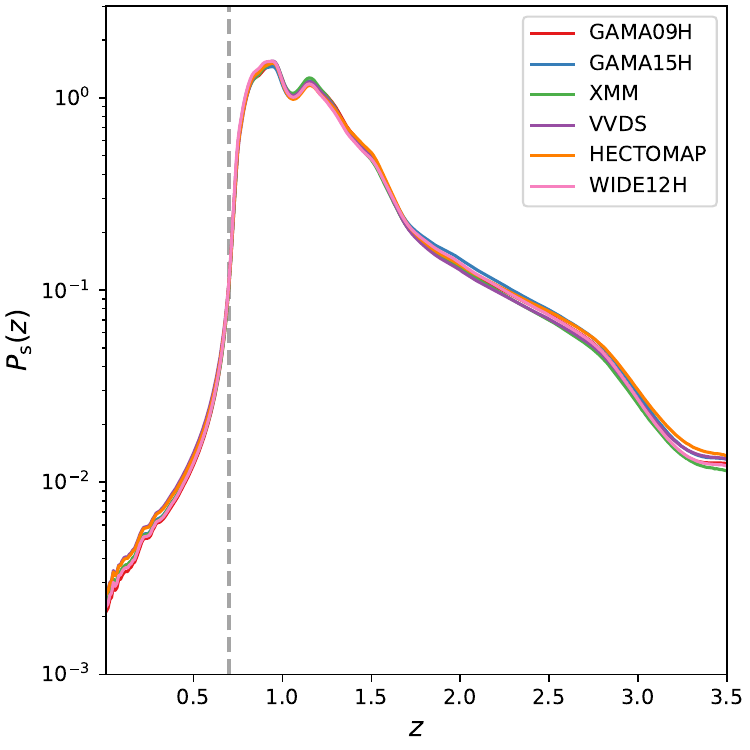}
}
\caption{
The stacked redshift distributions of the source samples.
The redshift distributions of individual subfields are indicated by different colors shown in the upper-right corner.
The vertical dash line marks the redshift of $\redshift = 0.7$, which approximately defines the threshold used in the source selection, i.e., equation~(\ref{eq:source_selection}). 
}
\label{fig:pz}
\end{figure}

We note that equation~(\ref{eq:calc_invsigmacrit}) is performed separately for the six subfields to account for the variation of the source redshift distributions.
We predict the dimensionless aperture mass peak \mkappa\ of a cluster using the stacked redshift distribution $P_{\mathrm{s}}\left(\redshift\right)$ of the subfield where the cluster is located.
The resulting stacked redshift distributions $P_{\mathrm{s}}\left(\redshift\right)$ are presented in Figure~\ref{fig:pz}, showing little variation among the six subfields.
The systematic uncertainty of the source redshift distribution is quantified in Section~\ref{sec:modelling_photoz_bias} and marginalized over in the analysis.

\subsection{Cluster abundance}
\label{sec:modelling_abundance}

After modelling the halo mass profiles (Section~\ref{sec:modelling_massprofile}) with the calibrated redshift distribution of the sources (Section~\ref{sec:modelling_Pz}), we now turn into the modelling of the observed cluster number.

The predicted number of shear-selected clusters detected with the signal-to-noise ratio \snr\ given a parameter vector $\vect{p}$ is related to 
the number of clusters with mass \mass\ at redshift \redshift,
$\frac{ \dif N\left(\mass,\redshift | \vect{p}\right) }{ \dif\mass\dif\redshift }$,
and the probability
$P\left(\snr | \mass, \redshift, \vect{p}\right)$, 
which characterizes the distribution of \snr\ given the cluster mass \mass\ at the redshift \redshift, as
\begin{equation}
\label{eq:nu-m-z-relation}
\frac{
\dif N\left(\snr | \vect{p} \right)
}{
\dif\snr
}
=
\int\dif\mass\int\dif\redshift
~
P\left(\snr | \mass, \redshift, \vect{p} \right) \frac{ \dif N\left(\mass,\redshift | \vect{p}\right) }{ \dif\mass\dif\redshift } 
\, ,
\end{equation}
in which
\begin{equation}
\frac{ \dif N\left(\mass,\redshift | \vect{p}\right) }{ \dif\mass\dif\redshift }
=
\frac{ \dif n\left(\mass, \redshift | \vect{p} \right) }{ \dif\mass }
\times \frac{ \dif V_{\mathrm{c}} \left(\redshift | \vect{p}\right) }{\dif \redshift} 
\times \Omega_{\mathrm{survey}} \, ,
\end{equation}
where $\frac{ \dif n\left(\mass, \redshift | \vect{p} \right) }{ \dif\mass }$ is the halo mass function,
$\frac{ \dif V_{\mathrm{c}}\left(\redshift | \vect{p}\right) }{ \dif \redshift }$ is the differential comoving volume at the redshift \redshift, and\
$\Omega_{\mathrm{survey}}$ is the area of the survey footprint.
The halo mass function is evaluated using the fitting formula calibrated against dark-matter-only simulations in \cite{bocquet16}. 
The parameter vector \textbf{p} contains all the free parameters in the modelling and its exact definition is given in Section~\ref{sec:modelling_statistics}.

The probability $P\left(\snr | \mass, \redshift, \vect{p} \right)$ is then expressed by an integral over the dimensionless aperture mass peak, \mkappa, and the angular size of the scale radius, \thetas, as
\begin{multline}
\label{eq:decomposition_of_intrinsic_and_measurement}
P\left(\snr | \mass, \redshift, \vect{p} \right) = \\
\int\int
P\left(\snr | \mkappa,\thetas  \right)
P\left(\mkappa,\thetas | \mass, \redshift, \vect{p} \right)
\dif\mkappa\dif\thetas
\, ,
\end{multline}
where $P\left(\mkappa,\thetas | \mass, \redshift, \vect{p} \right)$ describes the intrinsic scatter of \mkappa\ and \thetas\ at the fixed mass \mass\ and the redshift \redshift, and $P\left(\snr | \mkappa,\thetas  \right)$ accounts for the measurement uncertainty at the given \mkappa\ and \thetas.
By definition, for a cluster with a scale radius \rs\ at redshift \redshift, we have
\begin{equation}
\label{eq:thetas}
\thetas \equiv \frac{\rs}{D_{\mathrm{A}}\left(\redshift\right)} = \frac{\Rtwooo}{\ctwooo D_{\mathrm{A}}\left(\redshift\right) }\, .
\end{equation}

It is very important to stress that the predicted dimensionless aperture mass peak \mkappa\ differs from that of observed, which is denoted as \mkappanohat\ in equation~(\ref{eq:shear_field_convolution}).
The dimensionless aperture mass peak \mkappa\ is a theoretically predicted quantity assuming a spherically symmetric halo without miscentering, given a cluster at the redshift \redshift\ with the mass \mass, halo concentration \ctwooo, and the stacked redshift distribution $P_{\mathrm{s}}\left(\redshift\right)$.
Ideally, the observed and predicted mass peaks are expected to agree given the halo mass.
However, it is not true due to, e.g., the imperfect modelling of the cluster mass profile.
Given the cluster halo mass \mass, which is used to parameterize the halo mass function, the theoretically predicted \mkappa\ is therefore deviated from the observed aperture mass peak \mkappanohat.
The difference between \mass\ and \mwl\ is referred to as the ``weak-lensing mass bias'', which is taken into account in Section~\ref{sec:modelling_bwl_bias}.

We note that it is a natural choice to express the signal-to-noise ratio \snr\ in terms of \mkappa\ and \thetas\ in equation~(\ref{eq:decomposition_of_intrinsic_and_measurement}).
This is because the former and latter describe, respectively, the normalization and compactness of the cluster aperture mass profile, as the two most important factors determining the detectability of a source in an aperture-mass map.

The intrinsic scatter of \mkappa\ and \thetas\  at a fixed mass \mass\ and redshift \redshift, as described by $P\left(\mkappa,\thetas | \mass, \redshift, \vect{p} \right)$, is generally attributed to two causes.
The first is associated with the scatter in the halo concentration, resulting in the scatter in the angular size \thetas.
The second is attributed to the ``weak-lensing mass bias'' (see details in Section~\ref{sec:modelling_bwl_bias}) mainly arising from the inaccurate assumption about the mass profile, for which we assume a spherical NFW model in this work.
For example, the presence of correlated structures, halo triaxiality, and substructures leads to bias and scatter in the lensing aperture mass at a given halo mass \mass\ and redshift \redshift.
This can be characterized by a weak-lensing mass-to-mass-and-redshift (\mwl--\mass--\redshift) relation, which associates the weak-lensing mass \mwl\ with the halo mass \mass\ at the cluster redshift \redshift\ with intrinsic scatter.
Here, the weak-lensing mass \mwl\ is inferred to the mass that would have been obtained by modelling the weak-lensing observable using the chosen configuration (e.g., assuming a spherical NFW model) with an infinitesimal measurement uncertainty.
That is, the \mwl--\mass--\redshift\ relation with the intrinsic scatter accounts for the weak-lensing systematic uncertainty.
With this spirit, we rewrite the probability $P\left(\mkappa,\thetas | \mass, \redshift, \vect{p} \right)$ as
\begin{multline}
\label{eq:decomposition_of_intrinsic}
P\left(\mkappa,\thetas | \mass, \redshift, \vect{p} \right) = \\
\int 
P\left(\mkappa,\thetas | \mwl, \redshift, \vect{p} \right)
P\left(\mwl | \mass, \redshift, \vect{p} \right)
\dif\mwl \, , 
\end{multline}
in which $P\left(\mwl | \mass, \redshift, \vect{p} \right)$ is characterized by the \mwl--\mass--\redshift\ relation with the intrinsic scatter, 
and the probability $P\left(\mkappa,\thetas | \mwl, \redshift, \vect{p} \right)$ reads
\begin{multline}
\label{eq:concentration_scatter}
P\left(\mkappa,\thetas | \mwl, \redshift, \vect{p} \right) = \\
P\left(\mkappa |\thetas, \mwl, \redshift, \vect{p} \right)  
P\left(\thetas | \mwl, \redshift, \vect{p} \right) \, ,
\end{multline}
where $P\left(\thetas | \mwl, \redshift, \vect{p} \right)$ accounts for the intrinsic scatter of the halo concentration given the weak-lensing mass \mwl\ at the redshift \redshift, 
and $P\left(\mkappa |\thetas, \mwl, \redshift, \vect{p} \right)$ is a Dirac delta function centering at the value of the theoretically predicted aperture mass peak
given \mwl, \thetas, \redshift, and 
$P\left({\redshift}_{\mathrm{s}}\right)$ evaluated as equation~(\ref{eq:mkappa_field}).

The distribution $P\left(\thetas | \mwl, \redshift, \vect{p} \right)$ of the angular size is calculated using the distribution of the halo concentration \ctwooo, which is assumed to follow a log-normal distribution with the scatter \sigmaconcen\ and the mean value\footnote{
Note that the mean halo concentration is evaluated at the given weak-lensing mass \mwl\ instead of the halo mass \Mtwooo. 
This is consistent with typical weak-lensing studies, where the concentration is varied as a function of the inferred weak-lensing mass instead of the underlying halo mass.
}
predicted by the concentration-to-mass-and-redshift relation from \cite{diemer15} given the weak-lensing mass \mwl\ at the cluster redshift \redshift.

The characterization of the measurement uncertainty $P\left(\snr | \mkappa,\thetas  \right)$, as part of the calibration of the selection function, is described in Section~\ref{sec:modelling_selection_function} below.

In conclusion, with equations~(\ref{eq:nu-m-z-relation})~and~(\ref{eq:decomposition_of_intrinsic_and_measurement}) the differential number of clusters given a the signal-to-noise ratio \snr\ is
\begin{multline}
\label{eq:final_diff_n}
\frac{ \dif N\left(\snr | \vect{p} \right) }{ \dif\snr } =
\int\dif\mass\int\dif\redshift \frac{ \dif N\left(\mass,\redshift | \vect{p}\right) }{ \dif\mass\dif\redshift }  \times \\
\left[
\int\dif\mkappa\int\dif\thetas
P\left(\snr | \mkappa,\thetas  \right)
P\left(\mkappa,\thetas | \mass, \redshift, \vect{p} \right)
\right]
\, ,
\end{multline}
which includes the scatter in the halo concentration,
the measurement uncertainty in determining \snr\ (Section~\ref{sec:modelling_selection_function}), and 
the scatter and bias in the weak-lensing mass (Section~\ref{sec:modelling_bwl_bias}).

\subsection{The selection function}
\label{sec:modelling_selection_function}

The selection of the cluster sample is imposed as a cut on the signal-to-noise ratio \snr, i.e., $\snr \geq \numin$.
In what follows, we demonstrate that the calibration of the selection function in this work is to determine (1) the sample completeness $\comp\left(\mkappa,\thetas \right)$ and (2) the distribution of the signal-to-noise ratio \snr\ given the dimensionless aperture mass peak \mkappa\ and the angular size \thetas\ \textit{with} the presence of the sample cut $\snr \geq \numin$.
We note again that \mkappa\ is a quantity theoretically predicted at the weak-lensing mass \mwl, the angular size \thetas, and the cluster redshift given the distribution of source redshifts (see also Section~\ref{sec:modelling_abundance}).

\subsubsection{The formularism of the selection function}
\label{sec:selection_formula}

Consider a sample chosen with a
cut $\snr \geq \numin$.
The total number of clusters detected in the HSC survey reads
\begin{equation}
\label{eq:total_number_clusters}
\Nclusters = \Theta\left(\snr - \numin\right) \int \dif\snr 
\frac{ \dif N\left(\snr | \vect{p} \right) }{ \dif\snr } \, ,
\end{equation}
where $\Theta\left(\snr - \numin\right)$ is the Heaviside step function.
Substituting equation~(\ref{eq:final_diff_n}) into equation~(\ref{eq:total_number_clusters}), we express the probability $P\left(\snr | \mkappa,\thetas  \right)$ by combining with the selection $\Theta\left(\snr - \numin\right)$ as
\begin{eqnarray}
\label{eq:selection_function_rewrite}
\Theta\left(\snr - \numin\right)P\left(\snr | \mkappa,\thetas  \right) 
&\equiv &P\left(\snr, \snr \geq \numin| \mkappa,\thetas  \right)               \, \nonumber \\
&=      &P\left(\snr | \snr \geq \numin, \mkappa,\thetas  \right) \times       \, \nonumber \\
&       &P\left(\snr \geq \numin |  \mkappa,\thetas  \right)                   \, \nonumber \\
&\equiv &P_{\numin}\left(\snr | \mkappa,\thetas  \right) \times 			 \, \nonumber \\
&       &\comp\left(\mkappa,\thetas \right)                                 \, ,
\end{eqnarray}
where
the completeness at the given \mkappa\ and \thetas\ has the definition of
\begin{equation}
\label{eq:completeness_def}
\comp\left(\mkappa,\thetas \right) \equiv P\left(\snr \geq \numin |  \mkappa,\thetas  \right) \, ,
\end{equation}
and the distribution of the observed \snr\ at the given \mkappa\ and \thetas\ with the presence of the selection ($\snr \geq \numin$) is defined as
\[
P_{\numin}\left(\snr | \mkappa,\thetas  \right) \equiv P\left(\snr | \snr \geq \numin, \mkappa,\thetas  \right) \, .
\]
The second equality in equation~(\ref{eq:selection_function_rewrite}) holds according to the property of the conditional probability.
With equations~(\ref{eq:final_diff_n}) to (\ref{eq:selection_function_rewrite}), the total number of shear-selected clusters detected in the HSC survey reads
\begin{equation}
\label{eq:total_number_clusters_with_selection}
\Nclusters\left(\vect{p}\right) =  \int \dif\snr 
\frac{ \dif N_{\numin}\left(\snr | \vect{p} \right) }{ \dif\snr } \, ,
\end{equation}
where $\frac{ \dif N_{\numin}\left(\snr | \vect{p} \right) }{ \dif\snr }$ is the differential number of shear-selected clusters \textit{with} the presence of the sample selection,
\begin{multline}
\label{eq:diff_n_with_selection}
\frac{ \dif N_{\numin}\left(\snr | \vect{p} \right) }{ \dif\snr } \equiv
\Theta\left(\snr - \numin\right)\frac{ \dif N\left(\snr | \vect{p} \right) }{ \dif\snr }
 = \\
 \int\dif\mass\int\dif\redshift~
 \frac{ \dif N\left(\mass,\redshift | \vect{p}\right) }{ \dif\mass\dif\redshift }   \times \\
 \left[
\int\dif\mkappa\int\dif\thetas~
\comp\left(\mkappa,\thetas\right) P_{\numin}\left(\snr | \mkappa,\thetas  \right)
P\left(\mkappa,\thetas | \mass, \redshift, \vect{p} \right)
\right] 
\, .
\end{multline}

The remaining factors in equation~(\ref{eq:diff_n_with_selection}) to be determined are the probability $P_{\numin}\left(\snr | \mkappa,\thetas  \right)$ and the completeness $\comp\left(\mkappa,\thetas \right)$.
The distribution $P_{\numin}\left(\snr | \mkappa,\thetas  \right)$ 
is subject to observational effects, e.g., the variation of the survey depth, the shape measurement of the lensing sources, and the cosmic noise from large-scale structures.
In this work, we refer to these observational uncertainties as the measurement uncertainty.
In this way, at the given \mkappa\ and \thetas, the probability $P_{\numin}\left(\snr | \mkappa,\thetas  \right)$ accounts for the measurement uncertainty in the observed \snr, while $\comp\left(\mkappa,\thetas \right)$ describes the sample completeness with the presence of the measurement uncertainty and the selection $\Theta\left(\snr - \numin\right)$.

It is worth mentioning that the probability $P_{\numin}\left(\snr | \mkappa,\thetas  \right)$ differs from the probability $P\left(\snr | \mkappa,\thetas \right)$ in equation~(\ref{eq:decomposition_of_intrinsic_and_measurement}), where the latter does not include the selection function.
More explicitly, for $\snr > \numin$ the probability $P_{\numin}\left(\snr | \mkappa,\thetas  \right)$ is normalized and related to $P\left(\snr | \mkappa,\thetas  \right)$ as
\begin{equation}
\label{eq:normalize_Pmkappa}
P_{\numin}\left(\snr | \mkappa,\thetas  \right) 
= \frac{ \Theta\left(\snr - \numin\right) P\left(\snr | \mkappa,\thetas \right) }{ \comp\left(\mkappa,\thetas \right) }  \, .
\end{equation}

The determination of $P_{\numin}\left(\snr | \mkappa,\thetas  \right)$ and $\comp\left(\mkappa,\thetas \right)$ is carried out through an injection-based method, as described below.

\subsubsection{The injection-based approach}
\label{sec:injection}

The probability $P_{\numin}\left(\snr | \mkappa,\thetas  \right)$ and the completeness $\comp\left(\mkappa,\thetas \right)$ are quantified by injecting the shear signals of synthetic clusters into the real HSC weak-lensing aperture-mass maps.
This allows us to fully and self-consistently quantify the measurement uncertainty, including the noise from large-scale structures (i.e., cosmic shears).
This process is described and validated in depth in a companion paper \citep{chen24}, for which we refer readers to for more details.
In light of being self-contained in this paper, we provide a summary of the injection-based approach in Appendix~\ref{app:injection}.

In short, we generate millions of synthetic clusters spanning wide ranges of the halo mass, redshift, and the halo concentration, and inject their corresponding weak-lensing signals into the HSC-Y3 shape catalogs.
The injections are done on a per-halo basis, i.e., injecting one halo at a time, to avoid the blending between the injected halos.
After the injection, we run the peak-finding algorithm \citep[][see also Section~\ref{sec:clustersample}]{oguri21} to determine the weak-lensing signal-to-noise ratio \snr\ of the synthetic cluster as identical as what is done on the observed map.
For each injected cluster, we record the resulting \snr\ and the input \mkappa\ and \thetas.
By repeating this process for all synthetic clusters, we construct the sample completeness $\comp\left(\mkappa,\thetas \right)$ and the distribution $P_{\numin}\left(\snr | \mkappa,\thetas  \right)$.
In this approach, we naturally account for all observational systematics (e.g., masking and the depth variation) and measurement uncertainties (e.g., the shape noise and cosmic shears).

We emphasize that we inject the synthetic clusters over a larger footprint that completely covers the observed HSC-Y3 fields.
This is to account for the masking effect on the detectability of a weak-lensing peak, as a cluster could still create a peak in the observed aperture-mass maps even if the center is not located within the observed area or is masked by bright stars.
Specifically, we inject the clusters over areas $\Omega_{\mathrm{inj}}$ of $164.80$~deg$^{2}$, $59.28$~deg$^{2}$, $53.48$~deg$^{2}$, $204.25$~deg$^{2}$, $96.85$~deg$^{2}$, and $287.37$~deg$^{2}$ for the subfields of GAMA09H, GAMA15H, XXM, VVDS, HECTOMAP, and WIDE12H, respectively.
Consequently, the maximum of the completeness \comp\ derived from the injections does not reach unity but depends on the injected area $\Omega_{\mathrm{inj}}$.
Moreover, we have the effective survey area $\Omega_{\mathrm{survey}}$ as $\Omega_{\mathrm{survey}} \approx \Omega_{\mathrm{inj}} \times \mathtt{max}\left\lbrace \mathcal{C} \right\rbrace$.

\begin{figure*}
\centering
\resizebox{1.02\textwidth}{!}{
\includegraphics[scale=1]{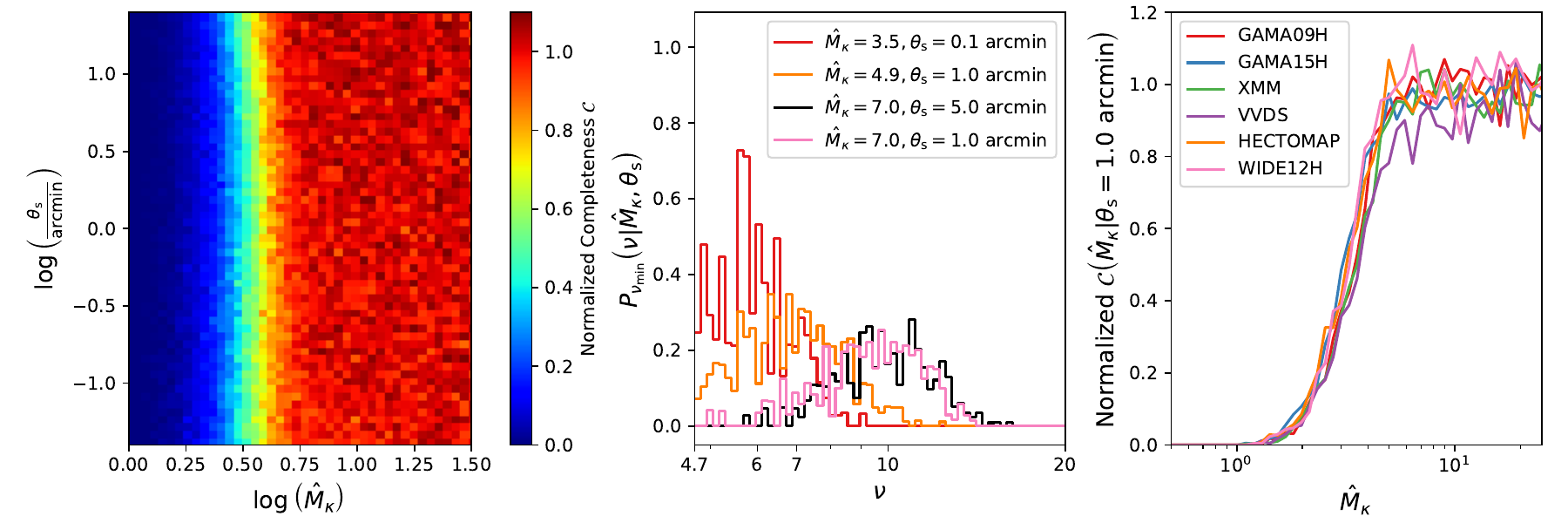}
}
\caption{
The selection function of the shear-selected clusters with the detection threshold $\numin=4.7$ and the selection parameter of $\deltasel\approx3$ (see text in Section~\ref{sec:injection}).
\textit{Left panel}:
The completeness $\comp\left(\mkappa,\thetas \right)$ as a function of the aperture mass \mkappa\ (the $x$-axis) and the angular scale \thetas\ in a unit of arcmin (the $y$-axis).
The level of the completeness at a given $\left(\mkappa, \thetas\right)$ is shown by the colorbar.
\textit{Middle panel}:
The probability $P_{\numin}\left(\snr | \mkappa,\thetas  \right)$, which describes the distribution of the signal-to-noise ratio \snr\ due to the measurement uncertainty at the given sets of $\left(\mkappa, \thetas\right)$. 
We show the results at $\left(\mkappa, \frac{ \thetas }{\mathrm{arcmin} }\right) = \left(3.5, 0.1\right), \left(4.9, 1.0\right), \left(7.0, 5.0\right),~\mathrm{and}~\left(7.0, 1.0\right)$ in red, orange, black, and pink, respectively.
\textit{Right panel}:
The normalized completeness as a function of \mkappa\ at $\thetas = 1~\mathrm{arcmin}$.
The results of the individual subfields are presented in different colors as indicated in the upper-left corner.
}
\label{fig:selection_function}
\end{figure*}

In practice, there are difficulties in matching a detected peak
to the injected clusters, especially at a low-\snr\ regime.
For example, an injected cluster with an intrinsic signal-to-noise ratio of ${\snr}_{\mathrm{int}}=1$ could be up-scattered to a final detection with $\snr=5$ if encountering an existing peak with ${\snr}_{\mathrm{cosmic}}=4$.
This is because we have $\snr\approx {\snr}_{\mathrm{int}} + {\snr}_{\mathrm{cosmic}}$ to the first-order approximation.
Similarly, an injected cluster with ${\snr}_{\mathrm{int}}=7$ could be down-scattered to the same \snr\ if encountering an existing void with ${\snr}_{\mathrm{cosmic}}=-2$.
In an extreme and unphysical case with an injected cluster with ${\snr}_{\mathrm{int}}=0$, the resulting distribution of \snr\ is just what we observe (${\snr}_{\mathrm{cosmic}}$) on the aperture-mass map.
It is then challenging to determine the sample completeness \comp\ at the low-\mkappa\ (i.e., low-mass) end, where a small difference could result in a statistically meaningful discrepancy in the predicted total number of detected clusters owing to the fact that low-mass halos significantly outnumber high-mass ones.
Therefore, the pivotal question is to what degree do we allow an injected cluster with the signal-to-noise ratio ${\snr}_{\mathrm{int}}$ to be considered a detection with the resulting peak height \snr\ given the observed fluctuation of the aperture-mass maps. 
This issue could be addressed by including the richness of the optical counterparts in quantifying the selection function \citep[see in-depth investigations in][]{chen24}.
However, including the optical richness into the forward modelling is beyond the scope of this work, for which we defer to a future work.
In this work, we choose to parameterize the selection function by an additional parameter \deltasel\ that is marginalized in the modelling,
as described below.

In addition to the standard peak finding algorithm \citep{oguri21}, we introduce two criteria for an injected cluster to be considered a successful detection. 
First, the position $\vect{x}$ of the detected peak must be within a $5$-arcmin separation from that ($\vect{x}_{\mathrm{diff}}$) on the difference map, i.e., $|\vect{x} - \vect{x}_{\mathrm{diff}}| \leq 5~\mathrm{arcmin}$.
Both the detected peak and that on the difference map are identified as the closest peaks to the true center of the injected cluster.
The difference map is calculated as the signal-to-noise map (or \snr-map) that subtracts the pre-injection \snr-map from the post-injection \snr-map.
That is, both the cosmic shears and shape noise observed on the map are subtracted off from the numerator of equation~(\ref{eq:snr_field}), while the shape noise in the denominator is preserved.
Therefore, the difference map manifests the pure signal-to-noise ratio of an injected cluster with the presence of variations of the shape noise and survey masking but \textit{without} the signals from cosmic shears. 
Second, the detected peak with the signal-to-noise ratio \snr\ must satisfy $|\snr - {\snr}_{\mathrm{diff}}| \leq  \deltasel$, where 
${\snr}_{\mathrm{diff}}$ is the peak height on the difference map and \deltasel\ is a nuisance parameter that needs to be marginalized over in the modelling.
The former criterion ($|\vect{x} - \vect{x}_{\mathrm{diff}}| \leq 5~\mathrm{arcmin}$) ensures that the detected peak must be associated with the injected cluster, even though the signal \snr\ could be dominated by an existing peak (i.e., ${\snr}_{\mathrm{cosmic}} \gg {\snr}_{\mathrm{diff}}$).
The latter ($|\snr - {\snr}_{\mathrm{diff}}| \leq  \deltasel$) describes the scattering range ($\pm\deltasel$) of the final \snr\ given the signal-to-noise ratio ${\snr}_{\mathrm{diff}}$ on the difference map.
Let us consider a value of $\deltasel=3$.
In a limit of a Gaussian field, this means that we only consider the $\approx99.7\percent$ confidence level of the scattering range (i.e., ${\snr}_{\mathrm{diff}} \pm 3$) for an injected halo with a signal-to-noise ratio ${\snr}_{\mathrm{diff}}$.

We derive a set of selection functions using different \deltasel\ in a range of $2.0$ to $4.5$ with a step of $0.25$, and then interpolate and marginalize the resulting selection function $P_{\numin}\left(\snr | \mkappa, \thetas\right)$ and $\comp\left(\mkappa,\thetas \right)$ as a function of the nuisance parameter \deltasel\ in the modelling.
The range of the grids in \deltasel\ used to generate the selection functions is gauged by comparing the richness distribution of the \texttt{CAMIRA} \citep{oguri14,oguri18} optical counterparts  of the shear-selected clusters with that predicted by the halo mass function given the selection function.
The prediction of the richness distribution for the shear-selected clusters is described and quantified in detail in \cite{chen24}.
In a nut shell, for a given selection function derived with a value of \deltasel, we calculate the corresponding completeness in terms of the halo mass \mass\ and redshift \redshift.
Then, the predicted number $N\left(\mass,\redshift\right)$ of clusters as a function of \mass\ and \redshift\ is calculated with the halo mass function in an assumed cosmology.
Leveraging the richness-to-mass-and-redshift (\rich--\mass--\redshift) relation with the measured scatter from \cite{murata19}, the corresponding richness distribution is predicted as $N\left(\rich\right) = \int \int P\left(\rich | \mass,\redshift\right) N\left(\mass,\redshift\right) \dif \mass\dif\redshift$.
To minimize the cosmological dependence, we only compare the shape of the richness distributions by normalizing the high-\rich\ end ($\rich > 30$), i.e., $P\left(\rich\right) = N\left(\rich\right) / N\left(\rich > 30\right)$.
We find that a possible range for \deltasel\ is
from $\approx3$ to $\approx4$.
Exceeding this range, the shape of the predicted richness distribution significantly deviates from that of observed.
Conservatively, we set a uniform prior on \deltasel\ ranging from $2.0$ to $4.5$ in the modelling to avoid the pile-up of the posterior on the prior boundary.
In Section~\ref{sec:results}, we see that a typical value of \deltasel\ is about $\approx3$, as self-calibrated by the abundance of the shear-selected clusters with $\numin = 4.7$.

Figure~\ref{fig:selection_function} contains the results of the selection function, where we set $\deltasel \approx 3$ and normalize the completeness \comp\ by a constant for a better visualization and visualization only.
The completeness $\comp\left(\mkappa,\thetas\right)$ is normalized by a constant defined as the area fraction of the injected footprint $\Omega_{\mathrm{inj}}$ where we have source galaxies, calculated using the algorithm of \texttt{HEALPix} with $\mathtt{NSIDE}=2048$. 
In the left panel, where the normalized completeness $\comp\left(\mkappa,\thetas\right)$ is shown, we can see that $\comp\left(\mkappa,\thetas\right)$ is nearly independent of the angular size \thetas.
This suggests that the detectability of a cluster under the truncated isothermal filter is only subject to the dimensionless aperture mass \mkappa.
The same picture is implied in the middle panel, showing that the distribution $P_{\numin}\left(\snr |\mkappa,\thetas\right)$ is a strong function of \mkappa\ with nearly zero dependence on \thetas.
In the right panel, we show the normalized completeness $\comp\left(\mkappa,\thetas\right)$ from the six subfields, indicating
little or no variations.
This again demonstrates the uniformity of the HSC survey.

It is worth mentioning that we have tried to vary the cosmology to generate the mock clusters and find negligible difference in the resulting selection function \citep[see][]{chen24}.
This suggests that the selection function is dominated by the measurement uncertainty observed on the aperture-mass map but not the underlying cosmology, as expected in the formularism developed in Section~\ref{sec:selection_formula}.

\subsection{Weak-lensing mass bias}
\label{sec:modelling_bwl_bias}

The gravitational lensing is one of the most reliable ways to estimate the mass of galaxy clusters, because it directly probes the underlying total potential of halos without making any assumptions about the dynamical state.
However, the weak-lensing mass \mwl\ is a biased estimate with respect to the halo mass \mass\ due to reasons including, but not limited to, the calibration bias in the galaxy shape measurement, the bias in the source redshift calibration, the lack of knowledge about the true cluster center, and the heterogeneous distribution of matter around halos and along the line of sight.
Among all these causes, the most dominant systematic bias  arises from the imperfect knowledge of the mass distribution of galaxy clusters, including the baryonic feedback.

For example, the heterogeneity of the cluster morphology---such as the halo concentration \citep{oguri05,bahe12,klypin16,lehmann17}, triaxiality \citep{clowe04,corless07,hamana12,chiu18b,herbonnet22,zhang23}, orientation \citep{herbonnet19,wu22}, and the presence of substructures or merging events \citep{okabe08,lee23}---leads to the bias in \mwl\ because the adopted model (e.g., the spherical NFW model) does not provide an accurate description for the halo mass distribution.
Meanwhile, correlated neighboring structures \citep{oguri11,gruen15} and uncorrelated large-scale structures along the line of sight \citep{hoekstra03,hoekstra11} introduce a bias and scatter in \mwl\ with respect to \mass.
The weak-lensing mass \mwl\ is also subject to the miscentering of galaxy clusters \citep{zhang19,sommer21} and the contamination from cluster member galaxies \citep{applegate14,hoekstra15,varga19}.
Importantly, the inclusion of the baryonic feedback effect in weak-lensing analyses has been demonstrated to be necessary toward the sub-percent accuracy \citep{lee18,schneider20,grandis21,cromer22,euclidCollaboration24,broxterman24}.

The bias and scatter in the weak-lensing mass \mwl\ can be calibrated against realistic halos in cosmological simulations. 
This task had been realized in \cite{grandis21} using the Magneticum Pathfinder Simulations \citep[][Dolag et al. in preparation]{dolag16} and had been also applied to constrain cluster scaling relations \citep{chiu22} and cosmology \citep{chiu23}.
In short, the weak-lensing mass bias and scatter with respect to the halo mass \mass\ at the cluster redshift \redshift\ was calibrated in \cite{chiu22} following the methodology described in \cite{grandis21}.
The resulting calibration is described by the so-called ``weak-lensing mass-to-mass-and-redshift (\mwl-\mass-\redshift) relation'', stating that the weak-lensing mass bias
\begin{equation}
\label{eq:bwl_def}
\bwl \equiv \frac{\mwl}{\mass} \, 
\end{equation}
follows a log-normal distribution around a mean value predicted by the \mwl--\mass--\redshift\ relation,
\begin{multline}
\label{eq:wlmass2halomass}
\left\langle\ln \left( \bwl | \mass,\redshift \right) \right\rangle
=
\ln\Awl + \\
\Bwl\ln\left(\frac{\mass}{\mpiv}\right) + 
\gammawl\ln\left(\frac{1 + \redshift}{1 + \zpiv} \right)
\, ,
\end{multline}
with intrinsic scatter \sigmawl.

In equation~(\ref{eq:wlmass2halomass}), \Awl\ is the normalization of the weak-lensing mass bias at the pivotal mass \mpiv\ and redshift \zpiv, and the parameters \Bwl\ and \gammawl\ describe the power-law indices of the mass and redshift trends, respectively.
In \cite{chiu22} and the follow-up work \cite{chiu23}, these parameters were calibrated with the pivotal mass $\mpiv = 2\times10^{14}\Msunh$ and redshift $\zpiv = 0.6$ using the halo mass definition of \Mfiveoo, which is defined as the enclosed mass with the spherically average density being $500$ times of the critical density.

We stress that equation~(\ref{eq:wlmass2halomass}) was specifically calibrated for the HSC weak-lensing analysis of the X-ray selected clusters studied in \cite{chiu22}, accounting for the multiplicative bias in the shape measurement, the redshift distribution and photo-\redshift\ bias of the source sample, the cluster member contamination, and the miscentering effect.
It is important to note that the uncertainties of the parameters \Awl, \Bwl, \gammawl, and \sigmawl\ include the baryonic feedback effect by quantifying the difference between the calibrations using dark-matter-only and hydrodynamics simulations \citep{grandis21}.

In this work, we account for the weak-lensing mass bias and scatter based on the information provided by equation~(\ref{eq:wlmass2halomass}).
Specifically, we incorporate equation~(\ref{eq:wlmass2halomass}) in two ways, namely, the ``mass-redshift (\mass-\redshift) dependent mass bias'' and the ``constant mass bias'', as detailed below.

\subsubsection{The mass-redshift dependent mass bias}
\label{sec:mz_dependent_bias}

In principle, the calibration result, i.e., equation~(\ref{eq:wlmass2halomass}), cannot be directly utilized for the shear-selected clusters, because different analysis methods are used between \cite{chiu22} and this work.
For instance, in \cite{chiu22} they selected galaxies securely at redshift $\redshift>\zcl+0.2$ as the source sample for a cluster at redshift \zcl; in this work, the source sample is uniformly selected at redshift $\redshift\gtrsim0.7$ for all clusters (see Section~\ref{sec:source_selection}).
Consequently, the weak-lensing mass of the shear-selected clusters at low redshift, e.g., $\redshift\approx0.3$, is estimated using a source sample at a higher redshift ($\redshift\gtrsim0.7$) with a lower source density compared to those ($\redshift>0.5$) used in \cite{chiu22}.
However, this difference raised from the \zcl-dependent source selection is expected to be at a level of $\lesssim3\percent$ for $\zcl\lesssim0.7$ \citep[see Figure~5 in][]{chiu22} and, hence, is insignificant compared to the scatter of the weak-lensing mass bias \citep[][and see also equation~(\ref{eq:wlpriors_mz}) below]{chiu23}.

Despite the differences, we argue that equation~(\ref{eq:wlmass2halomass}) is still expected to provide a fairly accurate description of the weak-lensing mass bias for the shear-selected clusters, given the similarity between \cite{chiu22} and this work:
First, both \cite{chiu22} and this work use the same HSC-Y3 shape catalog and, hence, share the same systematic uncertainty raising from the shape measurement.
Second, the spherical NFW model is commonly used to model the cluster mass profile, suggesting that the weak-lensing mass bias due to the misfitting to the observed clusters is expected to be consistent between \cite{chiu22} and this work.
Third, the cluster inner regions 
are both discarded in the modelling in \cite{chiu22} and this work.
This suggests that the scatter \sigmawl\ in both studies is reduced in a similar manner.
In fact, the inclusion of the cluster inner regions significantly increases the scatter \sigmawl\ of the weak-lensing mass bias by 
$\approx50\percent$
from the modelling of $\rdd\gtrsim0.5\Mpch$ to $\rdd\gtrsim0.2\Mpch$ \citep[see Figure~10 in][]{grandis21}.
This is expected, because the diversity of the cluster cores results in the large variation of the density profile at small radii \citep{tasitsiomi04}.

In this approach of the \mass-\redshift\ dependent mass bias, we directly use equation~(\ref{eq:wlmass2halomass}) obtained in \cite{chiu22} to account for the weak-lensing mass bias of the shear-selected clusters.
To do so, we convert the pivotal mass $\mpiv = 2\times10^{14}\Msunh$ from the mass definition of \Mfiveoo\ to that in terms of \Mtwooo, which is adopted in this work, assuming the concentration-to-mass relation from \cite{ishiyama21}.
This gives the new pivotal mass $\mpiv = 2.89\times10^{14}\Msunh$
at the pivotal redshift $\zpiv = 0.6$.
That is, we have the identical constraints on $\left(\Awl, \Bwl, \gammawl, \sigmawl\right)$ as in \cite{chiu23},
\begin{equation}
\label{eq:wlpriors_mz}
\centering
\begin{array}{ccc}
\Awl       &=   &  0.903 \pm 0.030  \\
\Bwl       &=   & -0.057 \pm 0.022 \\
\gammawl   &=   & -0.474 \pm 0.062 \\
\sigmawl   &=   &  0.238 \pm 0.037 \, ,
\end{array}
\end{equation}
at the pivotal mass $\mpiv = 2.89\times10^{14}\Msunh$ and redshift $\zpiv = 0.6$.

Finally, we use the constraints of equation~(\ref{eq:wlpriors_mz}) as the Gaussian priors applied on the parameters $\left(\Awl, \Bwl, \gammawl, \deltawl\right)$ in the modelling of cosmology.

The intrinsic scatter of the weak-lensing mass at a fixed halo mass was determined as $\sigmawl = 0.238 \pm 0.037$ in \cite{chiu23}.
The source of the intrinsic scatter includes the heterogeneity of the cluster mass profile, correlated neighboring structures, and baryonic feedback effects \citep{grandis21}.
This value is consistent with those estimated from a previous study of \cite{becker11}, where they obtained $\sigmawl\approx0.207\pm0.005$ and $\sigmawl\approx0.22\pm0.01$ for halos with $\Mfiveoo\gtrsim2\times10^{14}\Msunh$ at $\redshift=0.25$ and $\Mfiveoo\gtrsim1.5\times10^{14}\Msunh$ at $\redshift=0.5$, respectively.
Removing the contribution of uncorrelated line-of-sight structures would reduce the intrinsic scatter to $\sigmawl\approx0.18$ \citep{becker11,chen20}.
Similar results were also suggested by numerical simulations in \citet[][where $\sigmawl$ decreases from $\approx0.36$ to $\approx0.21$ as \Mtwooo\ increases from $\approx10^{14}\Msun$ to $\gtrsim10^{14.8}\Msun$]{bahe12} and 
\citet[][$\sigmawl\approx0.225$ at $\Mtwooo\approx8\times10^{14}\Msun$]{sommer22}.

Ideally, the weak-lensing mass bias of the shear-selected clusters as a function of \mass\ and \redshift\ shall be calibrated against simulations that include an end-to-end systematics exclusively associated with this work.
This is beyond the scope of this paper, and we leave this task to a future study for improvements.

\subsubsection{The constant mass bias}
\label{sec:constant_mass_bias}

In this approach, we apply a constant weak-lensing mass bias and scatter for the sample of the shear-selected clusters by leveraging the existing simulation-calibrated \mwl--\mass--\redshift\ relation, i.e., equation~(\ref{eq:wlmass2halomass}). 
By doing so, we effectively approximate the shear-selected clusters as the sample with a characteristic mass at a common redshift, for which the selected population shares the same weak-lensing mass bias on average.
This is also in line with other studies in quantifying the average mass bias of galaxy clusters selected in X-rays or the mm wavelength  \citep{vonderlinden14b,hoekstra15,PlanckCollaboration2015b,medezinski18b,miyatake19}.

Our goal is to derive a constant weak-lensing mass bias for the shear-selected sample with an uncertainty to account for variations in \bwl\ among the individual systems.
To incorporate equation~(\ref{eq:wlmass2halomass}), we must know the halo mass \mass\ and redshift \redshift\ of the shear-selected clusters to assess the weak-lensing mass bias \bwl.
Nevertheless, the mass \mass\ and redshift \redshift\ cannot be estimated for shear-selected clusters from weak lensing alone unless external information is utilized.
To this end, we use the external information from the optical properties of the shear-selected clusters to estimate the mass \mass, redshift \redshift, and then the weak-lensing mass bias \bwl.
We briefly describe the procedure here and refer the readers to Appendix~\ref{app:derive_bwl} for details. 

In short, we identify the optical counterparts of the shear-selected clusters in the HSC S21A \texttt{CAMIRA} optical cluster catalog \citep{oguri14,oguri18}.
This gives us the access to the cluster redshift \redshift\ and optical richness \rich\ of the shear-selected clusters.
With an existing and calibrated richness-to-mass-and-redshift (\rich--\mass--\redshift) relation of \texttt{CAMIRA} clusters, we estimate the inferred halo mass \mass\ of the shear-selected clusters based on the cluster redshift and optical richness, including the intrinsic scatter of the richness at fixed halo mass.
Once obtaining the richness-inferred halo mass \mass, we estimate the weak-lensing mass \mwl\ and the corresponding bias \bwl\ for each shear-selected cluster using the simulation-calibrated \mwl--\mass--\redshift\ relation in equation~(\ref{eq:wlmass2halomass}).
Finally, we derive the average $\left\langle\bwl\right\rangle$ of the individual shear-selected clusters as the resulting constant weak-lensing mass bias. 
We adopt three independently calibrated \rich--\mass--\redshift\ relation of the \texttt{CAMIRA} clusters \citep{murata19,chiu20,chiu20b} and repeat the above-mentioned procedure.
The final constraint on the constant weak-lensing mass bias of the shear-selected cluster sample is obtained as the mean of the resulting $\left\langle\bwl\right\rangle$ calculated with the three \rich--\mass--\redshift\ relations, leading to
\begin{equation}
\label{eq:wlpriors_constant}
\Awl       =     0.99  \pm 0.05 \, .
\end{equation}

Based on this, we apply a Gaussian prior $\mathcal{N}\left(0.99, 0.05^2\right)$ on the parameter \Awl\ for the modelling of the weak-lensing mass bias in the approach of the ``constant mass bias''.
The parameters \Bwl\ and \gammawl\ are set to be zero.
In terms of the intrinsic scatter, the Gaussian prior $\mathcal{N}\left(0.238, 0.037^2\right)$ is applied on the parameter \sigmawl, as identical to the ``\mass-\redshift\ dependent'' approach in Section~\ref{sec:mz_dependent_bias}.

\begin{figure}
\centering
\resizebox{0.5\textwidth}{!}{
\includegraphics[scale=1]{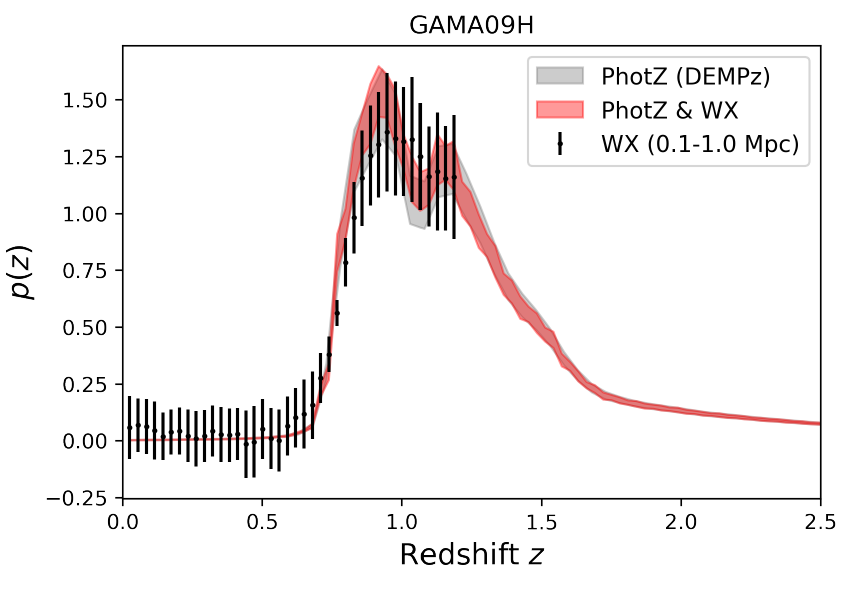}
}
\caption{
The calibration of the source redshift distribution.
The results of the GAMA09H field are shown, as an example.
The stacked redshift distribution $P_{\mathrm{s}}\left(\redshift\right)$ of the sources estimated by the code \texttt{DEmP} is in grey, while the independent calibration using the cross-correlation with the luminous red galaxies over the scale between $0.1~\Mpc$ and $1~\Mpc$ at $\redshift \leq 1.2$ is shown by the black points.
The joint calibration using the photometric and clustering-based estimates is in red.
An excellent consistency is seen, suggesting no significant bias in the redshift distribution of sources at $\redshift \leq 1.2$.
}
\label{fig:photoz_calib}
\end{figure}

\subsection{Photo-\redshift\ bias}
\label{sec:modelling_photoz_bias}

The stacked redshift distribution of the source sample is used to infer the critical surface mass density, i.e., equation~(\ref{eq:calc_invsigmacrit}).
Therefore, a bias in the source redshift distribution, if exists, would lead to a direct impact on the estimated aperture mass peak \mkappa\ and needs to be accounted for.

One of the most ideal ways to calibrate the redshift distribution of a photometrically selected source sample is to rely on their spectroscopic redshifts.
However, this task is challenging, or even unfeasible at this moment, given that the depth of the HSC imaging used to construct the shape catalogs is much deeper (with the limiting magnitude of $i\leq24.5$~mag)  than other existing spectroscopic observations.
In practice, previous HSC studies \citep[e.g.,][]{miyatake19,hikage19,hamana20,chiu22} had utilized the COSMOS 30-band photometry catalog \citep{ilbert09,laigle16} with extremely precise photo-\redshift\ to calibrate the selected sample of sources with a re-weighting technique \citep[see Section~7.1 in][]{nakajima12}, which accounts for the difference in the selection between the source and spectroscopic samples.
However, this method ultimately relies on the quality of the photo-\redshift\ in the COSMOS catalog, of which the outlier rate of the photo-\redshift\ at the faint end could post a concern on estimating the mean redshift of the source sample \citep[see discussions in][]{schrabback10,hildebrandt20}.
In addition, this approach assumes that the underlying galaxy population in the COSMOS field provides an unbiased and complete description of the source sample selected in the HSC survey.
This might not be true,
given the large cosmic variance at the scale of the COSMOS survey that is much smaller than the observed HSC footprint.

In this work, we address the potential photo-\redshift\ bias using two methods.
The first is referred to as the clustering-\redshift\ calibration, while the second is based on the photo-\redshift\ bias informed by the HSC-Y3 cosmic-shear analyses.
We describe them as follows.

\begin{figure}
\centering
\resizebox{0.47\textwidth}{!}{
\includegraphics[scale=1]{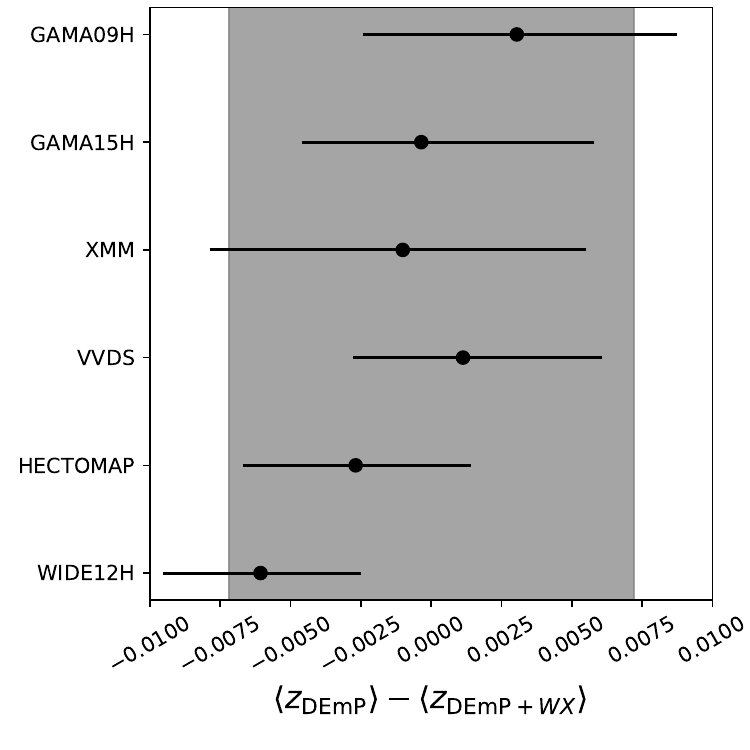}
}
\caption{
The estimated bias in the mean of the stacked redshift distribution in the subfields.
The bias \Deltaz\ is estimated as the difference in the mean redshifts between the stacked redshift distribution ($\left\langle\redshift_{\mathrm{DEmP}}\right\rangle$) and the joint clustering-and-photo-\redshift\ distribution ($\left\langle\redshift_{\mathrm{DEmP}+WX}\right\rangle$).
The quantity \Deltaz\ is defined as $\Deltaz \equiv \left\langle\redshift_{\mathrm{DEmP}}\right\rangle - \left\langle\redshift_{\mathrm{DEmP}+WX}\right\rangle$ in equation~(\ref{eq:delta_z_sys}).
A negative \Deltaz\ suggests that the photo-\redshift\ is biased low.
The grey area shows a range of $|\Deltaz| < 0.008$, which encloses the $\approx2.5$ times standard deviation of \Deltaz\ among the six subfields.
}
\label{fig:photoz_bias}
\end{figure}
\begin{figure*}
\centering
\resizebox{0.9\textwidth}{!}{
\includegraphics[scale=1]{
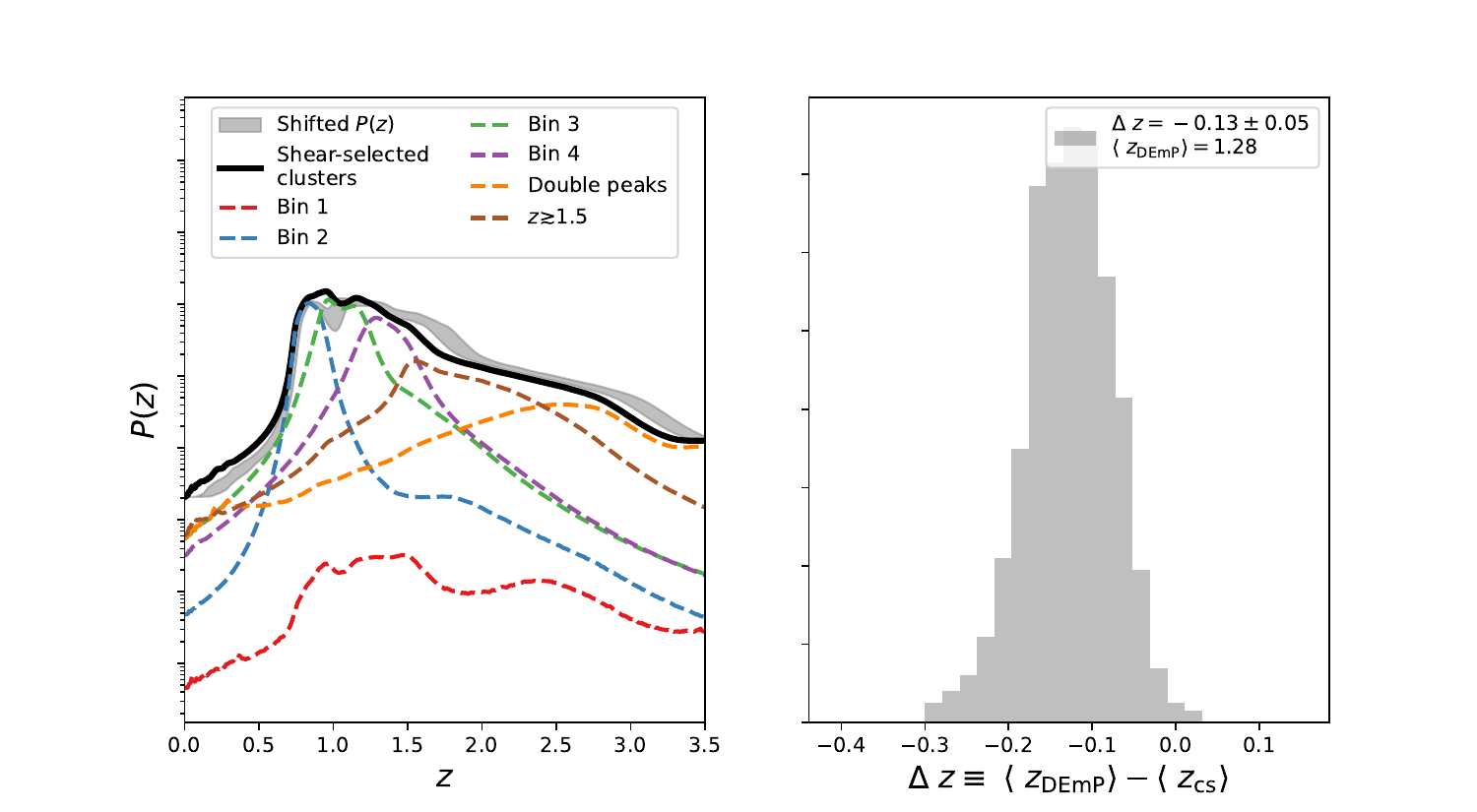
}}
\caption{
The calibration of the photo-\redshift\ bias informed by the HSC-Y3 cosmic-shear analyses.
\textit{Left panel}:
The stacked photo-\redshift\ distribution of the source sample (black curve) is decomposed into the four tomographic bins used in the HSC Y-3 cosmic-shear analyses ($\mathrm{Bin~1}$, $\mathrm{Bin~2}$, $\mathrm{Bin~3}$, and $\mathrm{Bin~4}$ in red, blue, green, and purple, respectively), and the remaining populations, including the high-\redshift\ sample (brown) and those with a double-peak feature in their photo-\redshift\ estimates (orange).
The $68\percent$ confidence level of the stacked redshift distribution including the photo-\redshift\ bias informed by the cosmic-shear analyses is shown as the shaded region. 
The $y$-axis is in an arbitrary unit at a logarithmic scale.
\textit{Right panel}:
The mean difference of the photo-\redshift\ bias parameter \Deltaz\ before and after including the photo-\redshift\ bias from the cosmic-shear analyses, i.e., equation~(\ref{eq:cs_z_bias}).
A negative \Deltaz\ suggests a biased-low photo-\redshift.
}
\label{fig:photoz_cs}
\end{figure*}

\subsubsection{The clustering-\redshift\ calibration}
\label{sec:clustering_deltaz}

To improve the redshift calibration in the latest HSC-Y3 analyses of cosmic shears \citep{li23,dalal23,more23,miyatake23,sugiyama23}, a new method to self-calibrate the redshift distribution of the selected source sample has been realized in \cite{rau23}.
In short, the redshift distribution of the HSC source sample is inferred using the angular cross-correlation 
with luminous red galaxies that have accurate redshift information (up to $\redshift\leq1.2$).
The cosmic variance arising from the limited footprint coverage of the HSC shape catalog is also taken into account.
The consistency of the redshift distributions between the photo-\redshift\ and the clustering-\redshift\ is quantified.
As a result, a conservative prior on the redshift distribution of the source sample is applied
to marginalize the systematic uncertainty of the photo-\redshift\ calibration.

In this work, we repeat the procedure in \cite{rau23} to calibrate the redshift distribution of the selected sources.
This is performed separately for the six subfields.
As an example, the results of the GAMA09H field are shown in Figure~\ref{fig:photoz_calib}.
As seen, the clustering-based redshift distribution (black points) shows excellent agreement with the photometrically derived stacked redshift distribution (grey area), suggesting no sign of bias in the photo-\redshift\ estimated by the code \texttt{DEmP}  at the overlapping redshift range $\redshift \leq 1.2$.
In addition, the joint clustering-and-photometry redshift distribution results in a more precise measurement (the red area).

We calculate the mean redshift inferred from the stacked photo-\redshift\ distribution of the sources (the grey area in Figure~\ref{fig:photoz_calib}) and that using the combined photometry and clustering technique (the red area in Figure~\ref{fig:photoz_calib}).
The difference between them, 
\begin{equation}
\label{eq:delta_z_sys}
\Delta\redshift \equiv 
\left\langle{\redshift}_{\mathrm{DEmP}}\right\rangle - 
\left\langle{\redshift}_{\mathrm{DEmP}+{W}_{X}}\right\rangle
 \, ,
\end{equation}
is attributed to the systematic uncertainty of the redshift calibration, 
where $\left\langle{\redshift}_{\mathrm{DEmP}}\right\rangle$ and 
$\left\langle{\redshift}_{\mathrm{DEmP}+{W}_{X}}\right\rangle$ are the mean redshifts inferred from the stacked $P\left(\redshift\right)$ and the joint method of photometry and clustering, respectively.
We have $\Delta\redshift \approx 0$ with scatter among the six subfields at a level of $0.0026$.
In Figure~\ref{fig:photoz_bias}, we show the results of the six subfields with the grey area indicating the range within the $\approx2.5$ times scatter, $|\Deltaz|<0.008$.

To account for the potential bias in the photo-\redshift\ calibration, we assume a systematic shift $\Delta\redshift$ in the stacked photo-\redshift\ distribution and apply a Gaussian prior $\mathcal{N}\left(0,0.008^2\right)$ on $\Delta\redshift$ to marginalize over the systematic uncertainty of the photo-\redshift.
Given the result, we effectively assume no photo-\redshift\ bias in this approach (the clustering-\redshift\ calibration).

\subsubsection{The photo-\redshift\ bias informed by cosmic shears}
\label{sec:cosmic_shear_deltaz}

The clustering-\redshift\ calibration mentioned above only utilizes the sample of luminous red galaxies out to $\redshift=1.2$, therefore such a calibration at $\redshift>1.2$ is not available.
In fact, the latest HSC-Y3 analyses of cosmic shears suggest a potential photo-\redshift\ bias 
at redshift $0.9\lesssim\redshift\lesssim1.5$ at a level of $2\sigma$ \citep[e.g., see Figure~12 in][]{dalal23}.
Assuming that such a photo-\redshift\ bias exists, we can estimate the potential photo-\redshift\ bias of our source sample using the information from the HSC-Y3 cosmic-shear analyses.
In the following, we describe the procedure to derive such a cosmic-shear (CS) informed photo-\redshift\ bias of our source sample.

Based on the information of the photo-\redshift,
the sources in the HSC-Y3 cosmic-shear analyses are divided into 
four tomographic bins, namely Bin~1, Bin~2, Bin~3, and Bin~4 at redshifts of 
$0.3\lesssim\redshift\lesssim0.6$, 
$0.6\lesssim\redshift\lesssim0.9$, 
$0.9\lesssim\redshift\lesssim1.2$,
$1.2\lesssim\redshift\lesssim1.5$, respectively.
The photo-\redshift\ bias of Bin~$i$ is then characterized by the parameter $\Delta\redshift_i$ (and $i=1,2,3,4$).
In their analyses, the photo-\redshift\ bias of the first two bins ($\Delta\redshift_1$ and $\Delta\redshift_2$) was marginalized over the Gaussian priors informed by the clustering-\redshift\ calibration \citep{rau23}.
Meanwhile, the last two bins (Bin~3 and Bin~4) were let free with flat priors, given the lack of available high-\redshift\ luminous red galaxies to carry out the clustering-\redshift\ calibration.
To explain the observed cosmic shears, they found that the posteriors reveal 
$\Delta\redshift_3\approx-0.12\pm0.05$ and
$\Delta\redshift_4\approx-0.19\pm0.08$, 
suggesting that the photo-\redshift\ estimates at 
$0.9\lesssim\redshift\lesssim1.2$ (Bin~3) and 
$1.2\lesssim\redshift\lesssim1.5$ (Bin~4) are biased low by 
$\approx0.12$ and $\approx0.19$, respectively.

Next, we cross-match the sample of our source galaxies with those used in the cosmic-shear analyses (via the unique $\texttt{object\_id}$ in the HSC catalog), and decompose our source sample into tomographic bins, including those used in the HSC-Y3 analyses.
As a result, we find that our source sample is composed of the HSC-Y3 tomographic 
Bin~1, Bin~2, Bin~3, and Bin~4 
at levels of 
$\lesssim0.05\percent$,
$\approx19.6\percent$,
$\approx39.3\percent$, and
$\approx23.1\percent$, respectively.
The remaining $\approx17.9\percent$ of our source galaxies are either attributed to the populations at redshift higher than those used in the HSC-Y3 cosmic-shear analyses ($\redshift\gtrsim1.5$; $\approx12.1\percent$), or identified as those with a double-peak feature in their photometric redshift distributions (see \citealt{rau23}; $\approx5.8\percent$).
These fractions are calculated including the lensing weight and show little variations among the subfields.

To estimate the photo-\redshift\ bias implied from the cosmic-shear analyses, we take the photo-\redshift\ bias $\left( \Delta\redshift_1, \Delta\redshift_2, \Delta\redshift_3, \Delta\redshift_4 \right)$ from the HSC-Y3 chains, then shift the photometric redshift distribution $p\left(\redshift\right)$ of individual source galaxies by an amount according to which bins they are located at after the decomposition, and finally estimate the CS-informed and lensing-weight weighted stacked redshift distribution $P\left(\redshift\right)$ following equation~(\ref{eq:wPz}).
For the source galaxies not in the HSC-Y3 tomographic bins, which occupy $\approx17.9\percent$ of the total populations, we apply the same shift $\Delta\redshift_4$ as in Bin~4, as a conservative approach.
The resulting tacked redshift distribution is referred to as the CS-informed redshift distribution $P_{\mathrm{cs}}\left(\redshift\right)$.
The photo-\redshift\ bias $\Delta\redshift$ given a set of $\left( \Delta\redshift_1, \Delta\redshift_2, \Delta\redshift_3, \Delta\redshift_4 \right)$ is determined as 
\begin{equation}
\label{eq:cs_z_bias}
\Delta\redshift \equiv 
\left\langle \redshift_{\mathrm{DEmP}}\right\rangle  -
\left\langle \redshift_{\mathrm{cs}}\right\rangle
\, ,
\end{equation}
where we define $\left\langle \redshift_{\mathrm{cs}}\right\rangle \equiv \int \redshift_{\mathrm{cs}} P_{\mathrm{cs}}\left(\redshift \right) \dif\redshift$.
We repeat the calculation of equation~(\ref{eq:cs_z_bias}) by sampling the chains of $\left( \Delta\redshift_1, \Delta\redshift_2, \Delta\redshift_3, \Delta\redshift_4 \right)$, resulting in a distribution of $\Delta\redshift$ that can be well described by a Gaussian distribution $\mathcal{N}\left(-0.13,0.05^2\right)$.
This result is presented in Figure~\ref{fig:photoz_cs}.
In the left panel, we see that the sources 
not used in the cosmic-shear analyses are mainly
located at $\redshift\gtrsim1.5$.
In addition, essentially no sources in Bin~1 is used in our source sample, given the stringent selection.
The $68\percent$ confidence level of the stacked redshift distribution including the photo-\redshift\ bias informed by the cosmic-shear analyses is shown by the shaded region
in the left panel, while the distribution of the resulting photo-\redshift\ bias parameter \Deltaz\ is shown in the right panel.

We apply a Gaussian prior $\mathcal{N}\left(-0.13,0.05^2\right)$ on the shift $\Delta\redshift$ in the stacked photo-\redshift\ distribution to account for the possible photo-\redshift\ bias in the modelling of cosmology.
As opposed to the clustering-\redshift\ calibration with the prior of $\mathcal{N}\left(0,0.008^2\right)$ in Section~\ref{sec:clustering_deltaz}, where essentially no photo-\redshift\ bias is present, here the CS-informed approach implies a photo-\redshift\ bias at a level of $\approx2.5\sigma$.
Moreover, this photo-\redshift\ bias is attributed to the sources at redshift $\redshift\gtrsim1.2$.
Without a large spectroscopic sample at such a high redshift, we are not able to verify and further quantify the photo-\redshift\ bias.
In this work, we therefore gauge the systematic uncertainty from the photo-\redshift\ by assessing the difference between the ``clustering-\redshift'' and ``CS-informed'' approaches.

\begin{table}
\centering
\caption{
The priors used in the modelling.
The first column contains the names of the parameters, while the second columns present the priors.
The default analysis choices of modelling the weak-lensing mass and photo-\redshift\ bias are marked by $\dagger$.
}
\label{tab:priors}
\resizebox{0.48\textwidth}{!}{
\begin{tabular}{ccc}
\hline\hline
Parameter &\multicolumn{2}{c}{Prior} \\
\hline
\multicolumn{3}{c}{Cosmology} \\
\hline
\omegam      &\multicolumn{2}{c}{$\mathcal{U}(0.1,0.99)$}             \\
\omegab      &\multicolumn{2}{c}{$\mathcal{U}(0.03,0.07)$}           \\
\omegak      &\multicolumn{2}{c}{Fixed to $0$}                       \\
\sigmaeight  &\multicolumn{2}{c}{$\mathcal{U}(0.45,1.15)$}           \\
\ns          &\multicolumn{2}{c}{$\mathcal{U}(0.92,1.0)$}            \\
\hnow        &\multicolumn{2}{c}{$\mathcal{U}(0.5,0.9)$}             \\
\w           &\multicolumn{2}{c}{Fixed to $-1$ or $\mathcal{U}(-2.5,-1/3)$}        \\
\hline
\multicolumn{3}{c}{Selection function (Section~\ref{sec:injection})} \\
\hline
\deltasel\   &\multicolumn{2}{c}{ $\mathcal{U}(2.0, 4.5)$  }    \\
\hline
\multicolumn{3}{c}{Weak-lensing mass bias (Section~\ref{sec:modelling_bwl_bias})} \\
\hline
&\mass-\redshift\ dependent bias & Constant bias$^{\dagger}$ \\[3pt]
\Awl\         &$\mathcal{N}(0.903,0.03^2)$    &   $\mathcal{N}(0.99,0.05^2)$    \\
\Bwl\         &$\mathcal{N}(-0.057,0.022^2)$  &   Fixed to $0$    \\
\gammawl\     &$\mathcal{N}(-0.474,0.062^2)$  &   Fixed to $0$    \\
\sigmawl\     &\multicolumn{2}{c}{$\mathcal{N}(0.238,0.037^2)$}    \\
\hline
\multicolumn{3}{c}{Photo-\redshift\ bias (Section~\ref{sec:modelling_photoz_bias})} \\
\hline
&Clustering-\redshift\ & Cosmic-shear-informed$^{\dagger}$ \\[3pt]
\Deltaz         &$\mathcal{N}(0,0.008^2)$    &   $\mathcal{N}(-0.13,0.05^2)$    \\
\hline
\multicolumn{3}{c}{Concentration} \\
\hline
\sigmaconcen         &\multicolumn{2}{c}{$\mathcal{N}(0.3,0.1^2)$}    \\
\hline
\end{tabular}
}
\end{table}
\begin{figure*}
\centering
\resizebox{0.85\textwidth}{!}{
\includegraphics[scale=1]{
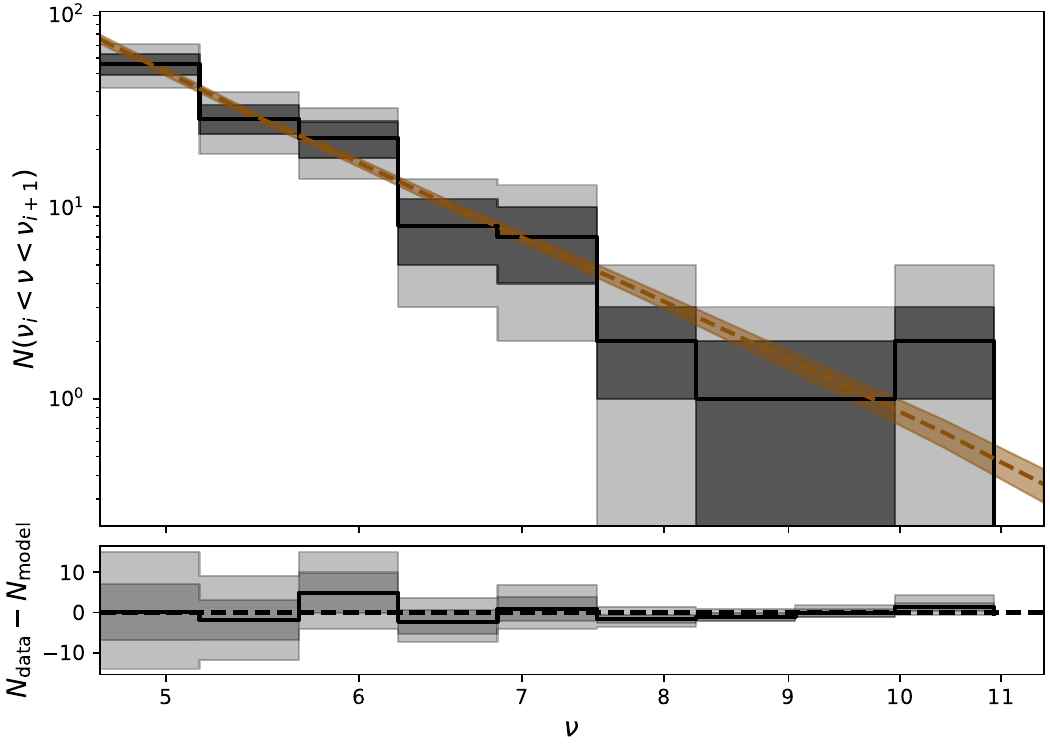
}}
\caption{
The comparison between the observed data and the best-fit model in the default analysis.
\textit{Upper panel}:
The total number of the shear-selected clusters with the selection of $\snr \geq \numin$ in the HSC-Y3 footprint, where $\numin = 4.7$.
We use $10$ equal-step logarithmic bins between $\nu = \numin$ and $\snr = 12$.
The per-bin number of the observed clusters is presented by the black line with the  Poissonian $68\percent$ and $95\percent$ confidence levels in the dark and light grey regions, respectively.
The best-fit model is plotted as the brown dashed line with the $68\percent$ confidence levels indicated by the shaded region.
\textit{Lower panel}:
The per-bin residual between the data $N_{\mathrm{data}}$ and the best-fit model $N_{\mathrm{model}}$ is shown by the black line with the $68\percent$ (dark-grey) and $95\percent$ (light-grey) confidence levels.
The perfect agreement is indicated by the dashed line.
The statistically consistency between the data and best-fit model is seen.
}
\label{fig:stacked_nu}
\end{figure*}

\subsection{Blinding}
\label{sec:blinding}

In this work, we carry out the analysis blindly to avoid the confirmation bias. 
The blinding is done on both the catalog and analysis levels, as described below.

On the catalog level, we blind the selection function, namely the completeness $\comp\left(\mkappa,\thetas\right)$ and the probability $P_{\numin}\left(\snr | \mkappa,\thetas  \right)$, which are derived using the injection method in Section~\ref{sec:injection}.
This is achieved by following the three-blinded-catalog approach that is widely used in the HSC weak-lensing analyses \citep[e.g.,][]{li23,dalal23}.
In short, we run the end-to-end analysis using the selection functions derived separately from three sets of blinded shape catalogs (labelled by $i$ with $i\in\left\lbrace1,2,3\right\rbrace$).
The three sets of the shape catalogs are identical except the multiplicative bias that is blinded as
\[
m_{\mathrm{blinded},i} = m_{\mathrm{true},i} + \dif m_{1,i} + \dif m_{2,i}  \, ,
\]
where $m_{\mathrm{blinded},i}$ and $m_{\mathrm{true},i}$ are the blinded and true multiplicative bias, respectively, and  $\dif m_{1,i}$ and $\dif m_{2,i}$ are two blinding terms.
The term $\dif m_{1,i}$ is known to the analysis team and must be subtracted from $m_{\mathrm{blinded},i}$ before any uses of the shape catalogs.
Meanwhile, the term $\dif m_{2,i}$ is kept encrypted to the analysis team and can only be decrypted by a blinded-in-chief who is not involved in the analysis.
The first blinding term $\dif m_{1,i}$ is non-zero and randomly sampled between $-0.1$ and $0.1$.
The second blinding term $\dif m_{2,i}$ is randomly chosen among $\left\lbrace-0.1,-0.05,0,0.05,0.1\right\rbrace$ such that $| \dif m_{2,i} - \dif m_{2,i + 1}| = 0.05$, where $i\in\left\lbrace 1, 2\right\rbrace$.
One and exactly one out of the three blinded shape catalogs has $\dif m_{2,i} = 0$, as the true shape catalog.
Consequently, in Section~\ref{sec:modelling_selection_function} the ellipticity of a synthetic cluster injected into the weak-lensing aperture-mass maps differs by a factor of $\approx\left(1 + \dif m_{2,j}\right)/\left(1 + \dif m_{2,i}\right)$ between the $i$-th and $j$-th blinded shape catalogs (see equation~(\ref{eq:cluster_injection_e})).
We repeat the injections for the three blinded catalogs, resulting in three sets of the blinded selection functions.
By design, the catalog blinding is achieved by carrying out the identical analysis on the three sets of the blinded selection functions separately.

On the analysis level, we do not reveal the absolute values of the cosmological parameters (i.e., \omegam, \sigmaeight, and $\seight\equiv\sigmaeight\left(\omegam/0.3\right)^{0.25}$) nor make any comparisons with other works before unblinding.

We unblind the results after 
(1) we pass the mock validation test that the exactly same code can recover the input values of the mock catalogs that are $10$ times larger than the data,  
(2) the parameter posteriors are converged and in agreement between different samplers (see Section~\ref{sec:modelling_statistics}),
(3) the posteriors of the cosmological parameters are not piling up at the edge of the flat priors,
(4) the constraints on \seight\ obtained from the three blinding catalogs behave reasonably with respect to the blinding factor $\left(1 + \dif m_{2} \right)$,
and
(5) the constraints on \omegam, \sigmaeight, and \seight\ are in excellent agreement (within $\lesssim0.2\sigma$) regardless the methods incorporating the weak-lensing mass bias (see Section~\ref{sec:modelling_bwl_bias}) and the photo-\redshift\ bias (see Section~\ref{sec:modelling_photoz_bias}).
No changes in the analysis are made after the unblinding.

\begin{figure*}
\centering
\resizebox{0.85\textwidth}{!}{
\includegraphics[scale=1]{
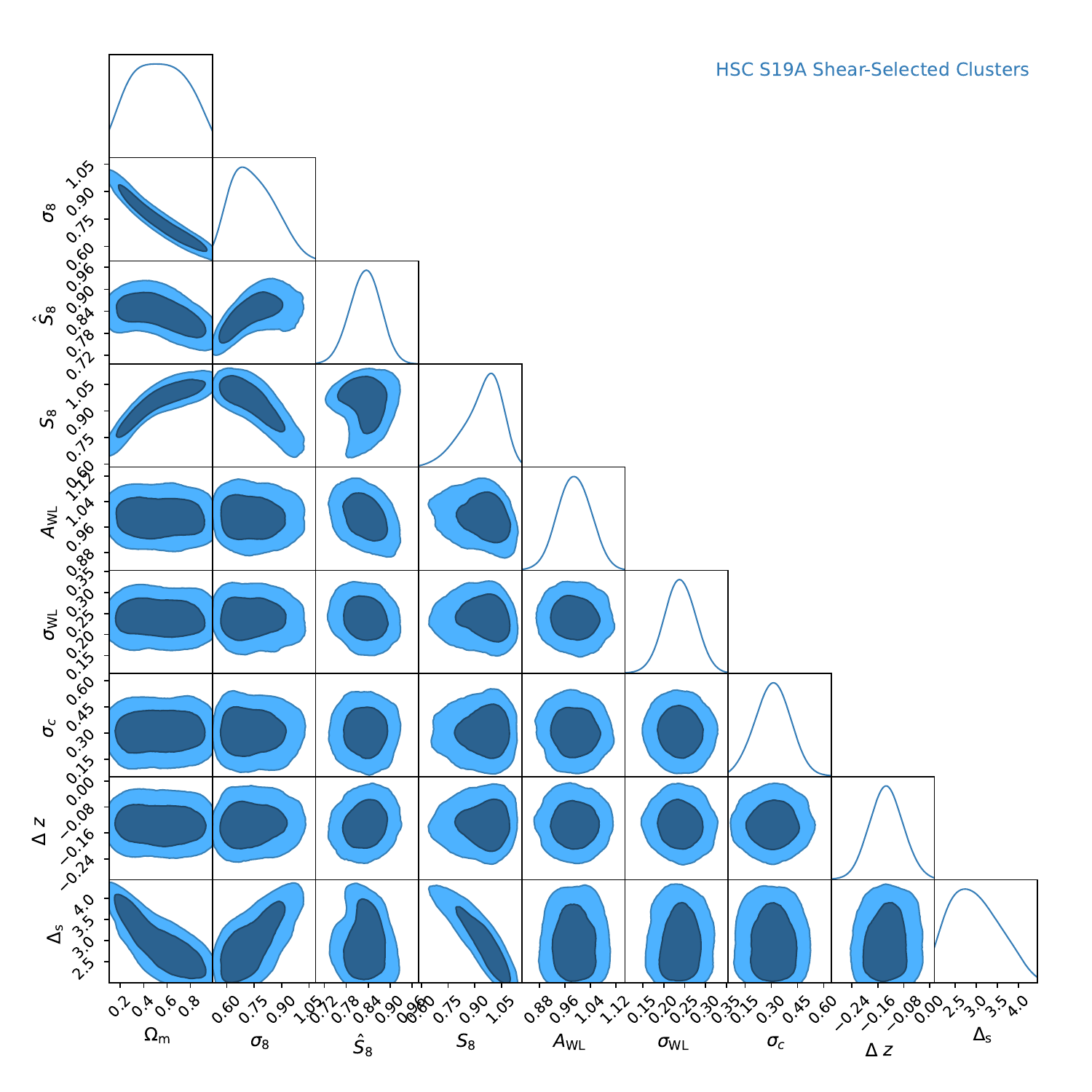
}}
\caption{
The posteriors (on-diagonal) of and covariances between (off-diagonal) the selected parameters in the default analysis.
In the off-diagonal plots, the $68\percent$ and $95\percent$ confidence levels between the parameters are indicated by the dark and light contours, respectively. 
This plot is made using the package \texttt{chainconsumer} \citep{hinton2016} with the option of $\mathtt{kde} = 1.5$ in the kernel density estimation.
}
\label{fig:gtc_som}
\end{figure*}
\begin{figure*}
\centering
\resizebox{0.9\textwidth}{!}{
\includegraphics[scale=1]{
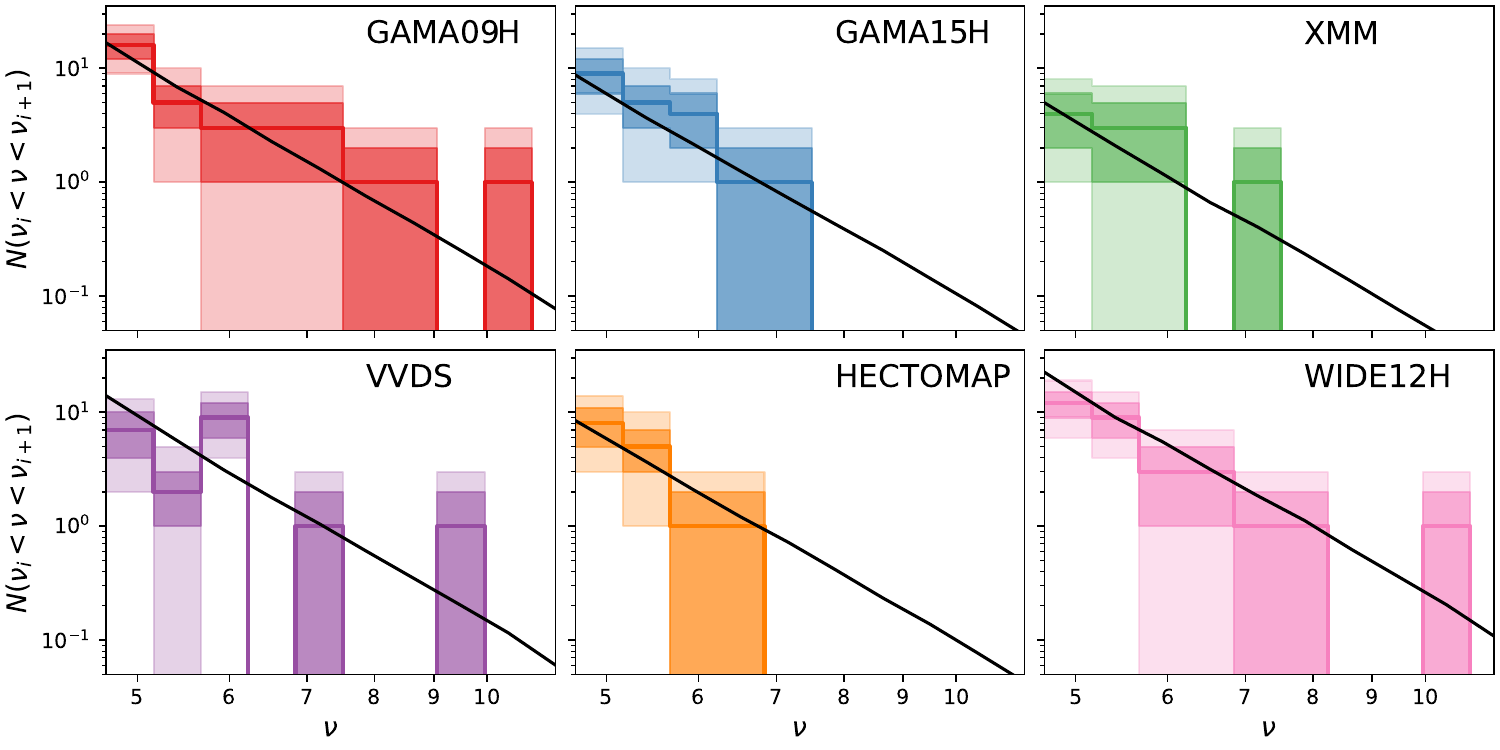
}}
\caption{
The number of the shear-selected clusters in the individual fields (indicated in the upper-right corner in each panel).
The same configuration in making the upper panel of Figure~\ref{fig:stacked_nu} is used.
No statistically significant discrepancy between the data and best-fit models is revealed.
}
\label{fig:subfield_nu}
\end{figure*}

\subsection{Statistical inference}
\label{sec:modelling_statistics}

We describe the statistical inference to constrain the cosmological parameters using the abundance of the shear-selected clusters in the HSC survey.
Given the observed data vector $\vect{D}$, which is the observed cluster number in each bin of the signal-to-noise ratio \snr, the posterior of the given parameter vector $\vect{p}$ is written as,
\begin{equation}
\label{eq:bayes_theorem}
P\left(\vect{p} | \vect{D}\right) 
\propto 
\mathcal{L}\left(\vect{D} | \vect{p}\right) \mathcal{P}\left( \vect{p} \right)
\, ,
\end{equation}
where $\mathcal{P}$ is the prior on the parameters $\vect{p}$, and $\mathcal{L}\left(\vect{D} | \vect{p}\right)$ is the likelihood of observing the data $\vect{D}$ 
given $\vect{p}$.
In this work, we adopt a Poisson likelihood to properly account for the small-number statistics \citep[see also the Cash statistics;][]{cash79}.
Namely, the log-likelihood of observing the number of the shear-selected clusters at the binning of \snr\ given the parameter vector $\vect{p}$ reads
\begin{equation}
\label{eq:poisson}
\ln\mathcal{L}\left(\vect{D} | \vect{p}\right) = \sum\limits_{i}
\Bigl( D_i \ln M_i \left(\vect{p}\right)  - M_i\left(\vect{p}\right) \Bigr)
\, ,
\end{equation}
where $i$ runs over the binning of \snr, 
$\vect{D}$ is the data vector composed of the observed number $D_i$ of the shear-selected clusters in the $i$th bin of $\snr_{i} < \snr \leq \snr_{i+1}$, 
i.e., $\vect{D}\equiv\left(D_1, D_2, \cdots\right)$, and
$M_i \left(\vect{p}\right)$ is the predicted number of clusters at the $i$th bin given the parameters $\vect{p}$, which is calculated using equation~(\ref{eq:diff_n_with_selection}) as
\begin{equation}
\label{eq:model_prediction}
M_i \left( \vect{p} \right)  = 
\int_{{\snr}_i}^{{\snr}_{i+1}} 
\frac{ \dif N_{\numin}\left(\snr | \vect{p} \right) }{ \dif\snr } \dif\snr
\, .
\end{equation}
In practice, equation~(\ref{eq:poisson}) is calculated separately for individual subfields and then summed up as a final log-likelihood.
For the data vector $\vect{D}$, we use $10$ logarithmic bins between $\numin = 4.7$ and $\snr = 12$.

In this work, the parameter vector \textbf{p} consists of
\begin{itemize}
\item The cosmological parameters: $\left\lbrace\omegam, \omegab, \sigmaeight, \ns, \hnow, \w\right\rbrace$,
\item The parameter to characterize the selection function: \deltasel ,
\item The parameters of the weak-lensing mass bias and scatter: $\left\lbrace\Awl,\Bwl,\gammawl,\sigmawl\right\rbrace$,
\item The parameter of the photo-\redshift\ bias: \Deltaz,
\item The scatter of the halo concentration: \sigmaconcen.
\end{itemize}
In what follows, we describe the priors $\mathcal{P}$ on these parameters.
The parameters and priors are summarized in Table~\ref{tab:priors}.

For the modelling of cosmology, we assume a flat \lcdm\ model characterized by the cosmological parameters of, at the present day, 
the mean matter fraction \omegam, 
the mean baryonic fraction \omegab, 
the r.m.s. of the density fluctuation \sigmaeight\ at a scale of $8\Mpch$,
the Hubble constant $h\equiv\Hnow/\left(100~\mathrm{km}/\mathrm{s}/\Mpc\right)$, and 
the spectral index \ns\ of the initial matter power spectrum.
We apply flat priors
$\mathcal{U}\left(0.1,0.9\right)$,
$\mathcal{U}\left(0.03,0.07\right)$,
$\mathcal{U}\left(0.45,1.15\right)$,
$\mathcal{U}\left(0.92,1.0\right)$, and
$\mathcal{U}\left(0.5,0.9\right)$ for
\omegam, \omegab, \sigmaeight, \ns, and $h$, respectively.
Meanwhile, we also consider a flat \wcdm\ cosmology with a flat prior $\mathcal{U}\left(-2.5,-1/3\right)$ on 
the parameter \w\ of the equation of state of dark energy.

To account for the scatter in the angular size of clusters (see Section~\ref{sec:modelling_abundance}), we allow an intrinsic scatter \sigmaconcen\ of the halo concentration-to-mass-and-redshift relation.
The Gaussian prior $\mathcal{N}\left(0.3, 0.1^2\right)$ is applied on \sigmaconcen.

The parameter \deltasel\ associated with the selection function is marginalized over a flat prior $\mathcal{U}\left(2.0,4.5\right)$, as described in Section~\ref{sec:injection}.

We conduct two approaches to account for the weak-lensing mass bias and scatter, namely the ``M-z dependent bias'' (Section~\ref{sec:mz_dependent_bias}) and the ``constant bias'' (Section~\ref{sec:constant_mass_bias}).
In the former, we include the parameters \Awl, \Bwl, \gammawl, and \sigmawl\ with the Gaussian priors of 
$\mathcal{N}(0.922,0.03^2)$, $\mathcal{N}(-0.057,0.022^2)$, $\mathcal{N}(-0.474,0.062^2)$, and $\mathcal{N}(0.238,0.037^2)$, respectively.
In the latter, we use the parameters of \Awl\ and \sigmawl\ to describe a global weak-lensing mass bias and scatter, on which the Gaussian priors of $\mathcal{N}(0.99,0.05^2)$ and $\mathcal{N}(0.238,0.037^2)$ are applied, respectively.

We introduce the parameter \Deltaz\ to account for the photo-\redshift\ bias for both the approaches of ``clustering-\redshift'' (Section~\ref{sec:clustering_deltaz}) and ``cosmic-shear'' (Section~\ref{sec:cosmic_shear_deltaz}) calibrations .
For the former we apply the Gaussian prior $\mathcal{N}\left(0, 0.008^2\right)$ on \Deltaz, while we adopt $\mathcal{N}\left(-0.13, 0.05^2\right)$ for the latter.

The exploration of the parameter space is carried out using the algorithm of nested sampling implemented in the code \texttt{Multinest} \citep{feroz08,feroz09,feroz19} that is run in the framework of \texttt{CosmoSIS} \citep{zuntz15}.

%
%

%
\begin{figure}
\centering
\resizebox{0.35\textwidth}{!}{
\includegraphics[scale=1]{
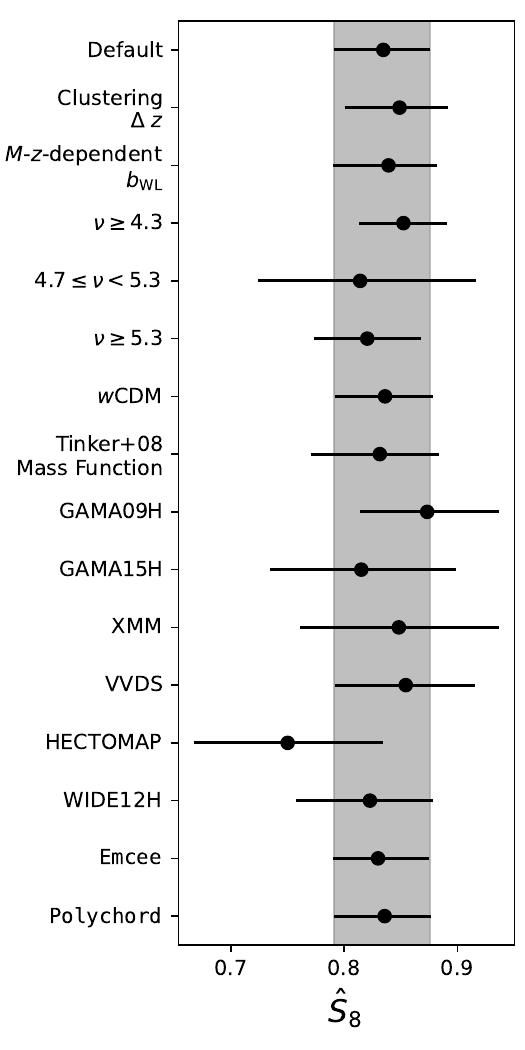
}}
\caption{
The systematic uncertainty in the marginalized posterior of \seight.
The result of the default analysis is on the top row with the $68\percent$ confidence level indicated by the vertical grey region.
The following rows represent the results from the different analysis choices made concerning the default analysis.
}
\label{fig:forest}
\end{figure}

\section{Results}
\label{sec:results}

To start with, in Section~\ref{sec:cosmological_constraints} we present our default analysis, which is the modelling of the abundance of shear-selected clusters with the selection $\snr \geq 4.7$ using the ``constant mass bias'' (Section~\ref{sec:constant_mass_bias}) and the ``CS-informed photo-\redshift\ bias'' (Section~\ref{sec:cosmic_shear_deltaz}).
Next, we assess the systematic uncertainties raised from the different analysis choices and sample selections in Section~\ref{sec:systematics}.
Finally, we compare ours cosmological constraints with external results in Section~\ref{sec:comparisons}.

\subsection{Cosmological constraints}
\label{sec:cosmological_constraints}

In Figure~\ref{fig:stacked_nu}, the upper panel shows the abundance of $129$ shear-selected clusters with the signal-to-noise ratio of $\snr \geq 4.7$ (the black steps) and the best-fit model (the brown dashed lines).
As seen, the residuals (the bottom panel) show no significant deviations from zero.
This suggests that the best-fit model provides an excellent description of the data.

Figure~\ref{fig:gtc_som} contains the parameter posteriors (the on-diagonal plots) and covariances (the off-diagonal plots) obtained in the default analysis.
The marginalized posteriors of the cosmological parameters are constrained as
\begin{eqnarray}
\label{eq:resulting_parameters}
\omegam                                                  & =  &  \omegamDefault                 \, ,       \\
\sigmaeight                                              & =  &  \sigmaeightDefault             \, ,       \\
\seight \equiv\sigmaeight\left(\omegam/0.3\right)^{0.25} & =  &  \seightDefault                 \, ,       \\
\seightnohat \equiv \sigmaeight\sqrt{ \omegam/0.3 }      & =  &  \seightNormDefault             \, ,  
\end{eqnarray}
while the other parameters of the weak-lensing mass bias, the photo-\redshift\ bias, and the concentration scatter are dominated by the Gaussian priors.
As a result, we put a constraint on \seight\ at a level of $\approx 5\percent$.
Note that the power-law index of \seight\ is chosen to be $0.25$ to minimize the degeneracy between \omegam\ and \sigmaeight.
Therefore, the parameter \seight\ represents the parameter space that is roughly perpendicular to the degeneracy direction between \omegam\ and \sigmaeight.

In this work, we cannot place meaningful constraints on the cosmological parameters of \omegam, \sigmaeight, and \seightnohat, except for \seight.
In addition, we observe strong degeneracy between the selection parameter \deltasel\ and the cosmological parameters of \omegam\ and \sigmaeight.
This is expected, because the selection parameter \deltasel\ dictates the amount of (up-)scattering from low-mass halos in the total number of cluster counts, for which we model it by the halo mass function that is primarily determined by both \omegam\ and \sigmaeight.
In other words, the cosmological constraints from the abundance $N\left(\snr\right)$ of shear-selected clusters are largely limited by the accuracy of the selection function or, more explicitly, the sample completeness \comp. 
As discussed in Section~\ref{sec:injection} and shown in the companion paper \citep{chen24}, it is feasible to include the modelling of the richness distribution $N\left(\rich\right)$ to tighten the constraint on \deltasel\ and then the cosmological parameters by breaking the degeneracy.
In this work, we choose to marginalize the cosmological constraints over the selection parameters \deltasel.

\begin{figure*}
\centering
\resizebox{0.48\textwidth}{!}{
\includegraphics[scale=1]{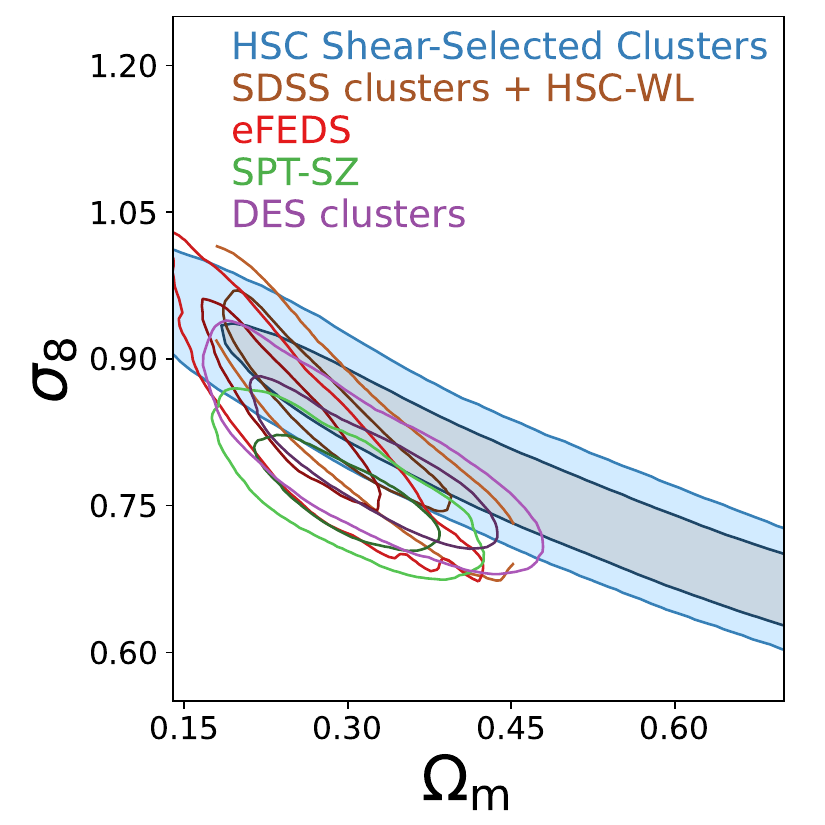}
}
\resizebox{0.48\textwidth}{!}{
\includegraphics[scale=1]{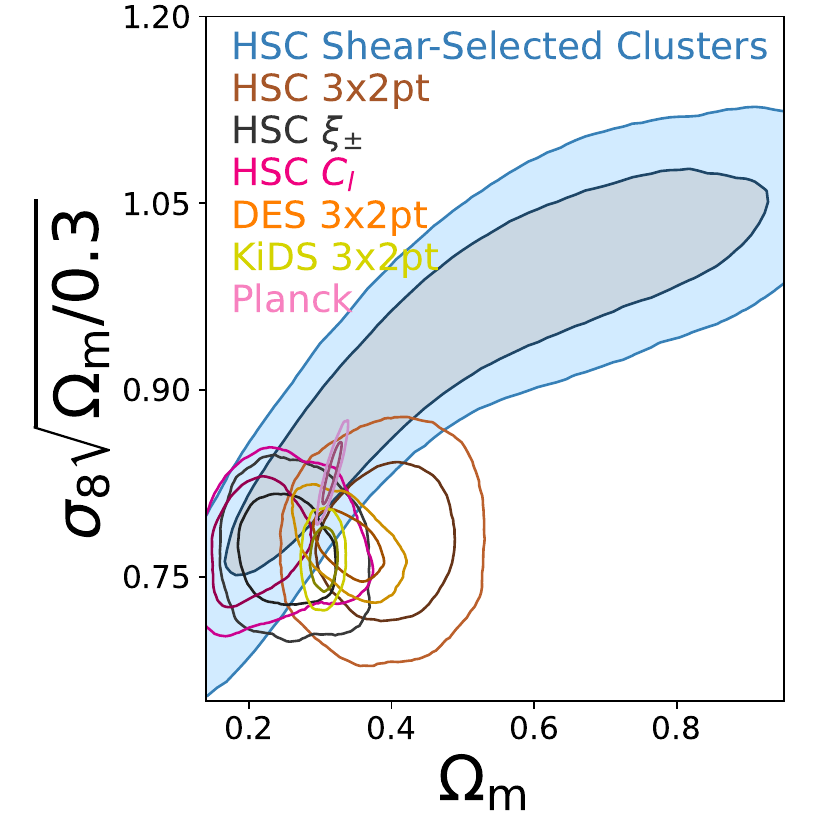}
}
\caption{
The comparisons of the cosmological parameters.
The contours represent $68\percent$ and $95\percent$ confidence levels of the parameter constraints.
\textit{Left panel}: The comparison in the \omegam-\sigmaeight\ space.
The constraints from the abundance of 
the HSC shear-selected clusters (this work), 
the SDSS optically selected clusters with the HSC weak-lensing mass calibration \citep[][]{sunayama24}, 
the eFEDS X-ray selected clusters \citep{chiu23},
the SPT SZ-selected clusters \citep{bocquet19}, and
the DES optically selected clusters \citep{costanzi21} are shown by 
the contours in
blue, brown, red, green, and purple, respectively.
For clarity, we only display the range of \omegam\ between $\approx0.15$ and $0.7$, despite the uniform prior $\mathcal{U}\left(0.1,0.99\right)$ being used.
\textit{Right panel}: The comparison in the \omegam-\seightnohat\ space.
The constraint from the abundance of the HSC shear-selected clusters is shown by the blue contours.
The results from the $3\times2$pt analyses of galaxy-galaxy lensing and clustering from
the HSC \citep{sugiyama23},
DES \citep{desy3kp-3x2}, and
KiDS \citep{heymans21} surveys are presented by the contours in
brown, orange, and yellow, respectively.
The constraints from the HSC-Y3 cosmic shears alone in the real \citep{li23} and fourier \citep{dalal23} spaces are 
shown by the black and bright pink contours, respectively.
The result from the \planck\ CMB analysis \citep[\texttt{TTTEEE-lowE}][]{PlanckCollaboration20} is color-coded by light pink. 
}
\label{fig:literature_comp}
\end{figure*}

It is worth mentioning that there exists mild degeneracy between \seight\ and the normalization \Awl\ of the weak-lensing mass bias. 
This suggests that the accuracy of \seight\ is directly related to the absolute calibration of the weak-lensing mass \mwl.
In this paper, the posterior of \Awl\ is dominated by the Gaussian prior that is informed by the existing simulation-based calibration of the weak-lensing mass bias \bwl. 
In the default analysis (using the ``constant mass bias''), the precision of the \bwl\ calibration is at a level of $0.05$, including baryonic feedback effects.
Clearly, a next step forward in a future study is to directly calibrate the weak-lensing mass bias of the shear-selected clusters using a dedicated simulation.

The modelling of the cluster abundance is carried out separately in the individual subfields and then stacked in the likelihood space to obtain the parameter posteriors.
In Figure~\ref{fig:subfield_nu}, we therefore show the predicted model of the cluster abundance in each subfield evaluated using the best-fit parameters.
As seen, no significant tension is present between the data and best-fit models. 
This suggests that the modelling is statistically consistent among the subfields.
We show this internal consistency test more quantitatively in a later paragraph.

\subsection{Systematics}
\label{sec:systematics}

From now on, we move to the examination of the systematic uncertainty due to
the different analysis choices and the selections.
We focus on the results of \seight, the parameter that we can put a statistically meaningful constraint on in this work.
For the comparisons of other cosmological parameters, we refer readers to Appendix~\ref{app:auxiliary}.
The systematic uncertainties in \seight\ are discussed as follows and visualized in Figure~\ref{fig:forest}.

Changing the modelling of the weak-lensing mass bias to the ``\mass-\redshift\ dependent'' approach (Section~\ref{sec:mz_dependent_bias}) results in a constraint as $\seight = \seightMZdep$,
which is fully consistent with the default result ($\seight = \seightDefault$ using the ``constant mass bias'').
Switching to the modelling of the photo-\redshift\ bias to the ``clustering-\redshift'' approach, which effectively implies $\Deltaz\approx0$,
delivers a constraint of $\seight = \seightClusterz$, which is consistent with the default result at a level of $\lesssim1\sigma$.
This suggests that the analysis choice made to account for the weak-lensing mass bias and the photo-\redshift\ bias is a subdominant factor.

As a test on the internal consistency, the modelling of the low-\snr\ ($4.7\leq\nu<5.3$; $63$ clusters) and high-\snr\ ($\snr\geq5.3$; $66$ clusters) subsamples leads to constraints as
$\seight = \seightLowSN$ and
$\seight = \seightHghSN$, respectively.
Lowering the selection threshold to $\numin = 4.3$ gives a constraint of $\seight = \seightLowNumin$ with $207$ shear-selected clusters.
In addition, fitting the cluster abundance of each single subfield alone gives 
a statistically consistent constraint on \seight\ 
with the default result at a level of $\lesssim1\sigma$, albeit with a much larger errorbar (see Figure~\ref{fig:forest}).
To conclude, we find no statistically significant difference of all these subsample modelling\footnote{We have re-run the cross-matching between these subsamples and the optical counterparts following the same procedure in Section~\ref{sec:constant_mass_bias}, 
and found negligible difference in the resulting constant mass bias \bwl.
Therefore, we use the same Gaussian prior $\mathcal{N}\left(0.99, 0.05^2\right)$ on \Awl\ for the modelling of these subsamples.} 
(low-\snr, high-\snr, low-\numin, and single-subfield) compared to the default analysis.
This strongly ensures that no internal tension is revealed in the data.

We test the analysis choice in evaluating the halo mass function, which is based on the \cite{bocquet16} fitting formula, by default. 
We repeat the analysis using the \cite{tinker08} mass function and find no significant difference, as seen in Figure~\ref{fig:forest}.
This suggests that our constraints are not subject to the choice of the fitting formula of the halo mass function.
Assuming a flat \wcdm\ model with a varying equation of state of dark energy \w, we obtain a fully consistent constraint $\seight = \seightWcdm$.
We also run different samplers \texttt{Emcee} \citep{foreman13,foreman19} and \texttt{Polychord} \citep{handley15}, and confirm that we obtain fully consistent results.

\begin{figure*}
\centering
\resizebox{0.50\textwidth}{!}{
\includegraphics[scale=1]{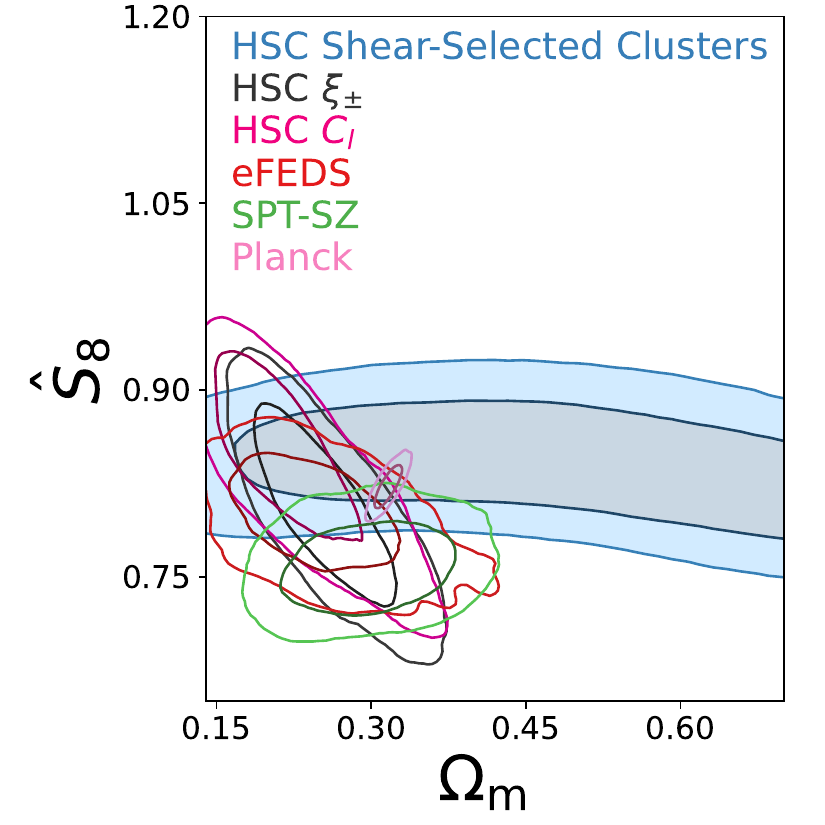}
}
\resizebox{0.45\textwidth}{!}{
\includegraphics[scale=1]{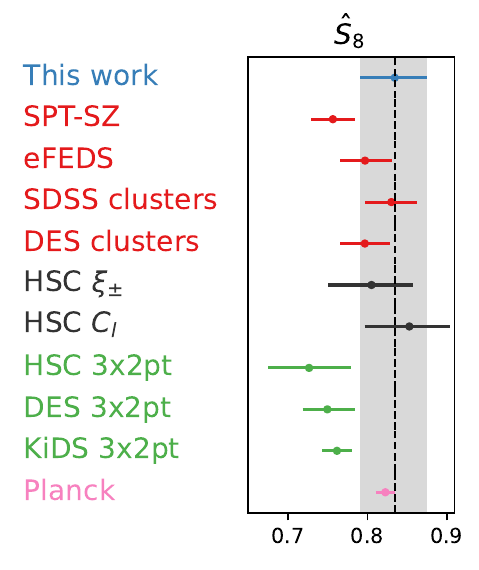}
}
\caption{
The comparisons of the cosmological parameters.
The contours represent $68\percent$ and $95\percent$ confidence levels of the parameter constraints.
\textit{Left panel}: The comparison in the \omegam-\seight\ space with the same color codes used as in Figure~\ref{fig:literature_comp}.
Similarly, only the range of \omegam\ between $\approx0.15$ and $0.7$ is shown for clarity.
\textit{Right panel}: The comparison of the marginalized posterior of \seight.
The result of this work is shown on the top row with the $68\percent$ confidence level indicated by the vertical grey region.
The constraints from the abundance of SPT-SZ \citep{bocquet19}, eFEDS \citep{chiu23}, SDSS \citep{sunayama24}, and DES \citep{costanzi21} clusters are in red.
The results of the cosmic shears alone from HSC-Y3 \citep{li23,dalal23} are in black.
The $3\times2$pt analyses of HSC-Y3 \citep{sugiyama23}, DES \citep{desy3kp-3x2}, and KiDS \citep{heymans21} are in green.
The CMB analysis from \planck\ \citep{PlanckCollaboration20} is shown in light pink.
Overall, excellent agreement is seen with other probes, except for the $3\times2$pt analyses with mild discrepancy at levels of $\approx2\sigma$.
}
\label{fig:literature_comp2}
\end{figure*}

\subsection{Comparisons with other studies}
\label{sec:comparisons}

In this section, we compare our constraint with results from the literature.
The consistency between our result and other studies is quantified by using the code \texttt{PosteriorAgreement}\footnote{\url{https://github.com/SebastianBocquet/PosteriorAgreement}}.

In the left panel of Figure~\ref{fig:literature_comp}, we show the comparisons in terms of the \omegam-\sigmaeight\ space with other cluster-abundance-based studies from 
the South Pole Telescope SZ survey \citep[SPT-SZ in green;][]{bocquet19}, 
the \erosita\ Final Equatorial-Depth Survey \citep[eFEDS in red;][]{chiu23},
the Dark Energy Survey \citep[DES in purple;][]{costanzi21},
and the optically selected clusters from the Sloan Digital Sky Survey (SDSS) survey with the HSC weak-lensing calibration \citep[SDSS in brown;][]{sunayama24}.
The result from the abundance of the shear-selected clusters reveals good agreement with those from the
SPT-SZ, eFEDS, DES, and SDSS clusters
at levels of 
$2.1\sigma$,
$0.9\sigma$,
$1.0\sigma$, and
$0.2\sigma$, respectively.
Despite the mild discrepancy seen in the SPT-SZ sample (green contours), it is worth mentioning that the updated analysis of the SPT-SZ clusters with the DES weak-lensing mass calibration preferred a higher value of \sigmaeight\ \citep[see Figure~5 in][]{bocquet24} and, hence, results in better agreement with ours.
Overall, the constraints from the abundance of the other cluster samples selected based on a baryonic tracer are all statistically consistent with our result---the only one among these whose selection
is completely free from any assumptions about the baryons physics.

Next, we show the comparisons in terms of the \seightnohat-\omegam\ space with the results of 
$3\times2$pt galaxy-galaxy lensing and clustering from 
HSC-Y3 \citep[in brown;][]{sugiyama23},  
DES \citep[in orange;][]{desy3kp-3x2}, and
Kilo-Degree Survey \citep[KiDS in yellow;][]{heymans21}
in the right panel of Figure~\ref{fig:literature_comp}.
We additionally show the HSC-Y3 constraints solely based on cosmic shears in the real \citep[black contours;][]{li23} and fourier \citep[bright pink contours;][]{dalal23} spaces.
The comparisons with the HSC-Y3 cosmic shears \citep{li23,dalal23} are interesting, given that the same weak-lensing data sets are used in this work.
In spite of the largely unconstrained parameter \omegam, our result shows a slightly higher value of \seight\ compared to all the $3\times2$pt analyses (HSC-Y3, DES, and KiDS).
Considering the degeneracy, the discrepancy with the $3\times2$pt analyses of HSC-Y3, DES, and KiDS is not statistically significant at levels of
$1.9\sigma$,
$2.0\sigma$, and
$2.5\sigma$, respectively.
Intriguingly, the cosmic-shear-alone analyses (black and bright pink contours), which also relied on the HSC-Y3 weak-lensing data, show good agreement with our results at levels of $\lesssim0.6\sigma$, despite the large errorbars in \omegam.
This suggests that (1) no internal inconsistency within the HSC-Y3 data sets is revealed between cosmic shears and the cluster abundance of shear-selected clusters, and that (2) the mild discrepancy seen with the HSC-Y3 $3\times2$pt could be attribute to the analysis of galaxy-galaxy clustering.

In the right panel of Figure~\ref{fig:literature_comp}, the constraint from Cosmic Microwave Background (\texttt{TTTEEE-lowE}) observed by \planck\ \citep[][]{PlanckCollaboration20} is shown in light pink.
Our result is in good agreement with \planck\ at a level of $0.8\sigma$.

Finally, we compare our results with others in terms of \seight, which represents the parameter space perpendicular to the \omegam-\sigmaeight\ degeneracy.
In the \omegam-\seight\ space, the comparisons with the HSC-Y3 cosmic shears, the abundance of eFEDS and SPT-SZ clusters, and the \planck\ results are presented in the left panel of Figure~\ref{fig:literature_comp2}.
It is seen that our result shows not only a consistent constraint but also an almost orthogonal degeneracy with respect to the results from the HSC-Y3 cosmic shears.
This suggests that a joint analysis of the cosmic shears and the cluster abundance could break the parameter degeneracy and improve the constraint obtained from the HSC-Y3 data.
Overall, the consistency with the abundance of eFEDS and SPT-SZ clusters is clearly seen.
Again, the updated analysis of SPT-SZ clusters delivers a constraint of $\seight=0.805\pm0.016$ \citep{bocquet24} that is in better agreement with ours.
Meanwhile, the agreement with the \planck\ result suggests no \seightnohat\ tension.

We summarize the comparisons in terms of the marginalized posterior of \seight\ in the right panel of Figure~\ref{fig:literature_comp2}, where the results based on the cluster abundance are in red, the analysis of HSC-Y3 cosmic shears alone are in black, the $3\times2$pt analyses are in green, and the \planck\ result is in light pink. 
As seen, our result ($\seight=\seightDefault$) show excellent agreement with the other probes, except for the $3\times2$pt analyses with the mild discrepancy at levels of $\approx2\sigma$.

%
%

\section{Conclusions}
\label{sec:conclusions}

In a series of two papers \citep[this work and][]{chen24}, we present a cosmological study using the sample of weak-lensing shear-selected clusters in the HSC survey.
The clusters are selected as peaks on the aperture-mass maps constructed using the latest HSC-Y3 weak-lensing data sets, which consist of six subfields spanning a total area of $\approx500~\mathrm{deg}^2$.
The selection of the cluster sample is purely based on the technique of weak gravitational lensing and is entirely independent of any baryonic tracers, leading to a gravity-only selection.
To avoid the weak-lensing contamination by cluster member galaxies, the aperture-mass maps are constructed using the source galaxies securely selected at redshift $\redshift\gtrsim0.7$ and convolved with an optimized kernel that excises the core of clusters.
This results in a sample of $129$ shear-selected clusters with a weak-lensing signal-to-noise ratio \snr\ of $\snr\geq\numin\equiv4.7$.
Through cross-matching with existing catalogs of optically selected clusters, primarily the \texttt{CAMIRA} catalog constructed using the HSC data across the same footprint, the shear-selected cluster sample spans a redshift range of $\redshift\lesssim0.7$ (median redshift $\approx0.34$) with a purity of $\gtrsim99\percent$.
This is by far the largest sample of shear-selected clusters used in a cosmological study, owing to the HSC survey that provides deep and uniform photometry that covers a wide area on the sky.

The cosmological constraints are obtained by forward-modelling the cluster abundance in terms of the cluster counts in \snr, $N\left(\snr\right)$.
To do so, we establish the relation between the observed weak-lensing signal \snr, which is the mass proxy of the shear-selected clusters, and the underlying halo mass \mass\ that is used to parameterize the halo mass function.
We develop a novel framework to model the \snr--\mass\ relation, which is composed of two components: the intrinsic distribution of the weak-lensing mass \mwl\ and the measurement uncertainty of the observed \snr.

The intrinsic distribution of \mwl\ is modelled by the weak-lensing mass bias $\bwl\equiv \mwl/\mass$ as a function of the cluster mass \mass\ at a given redshift.
We utilize an existing \bwl--\mass--\redshift\ relation calibrated against cosmological simulations to account for the bias and the intrinsic scatter in \mwl, which is attributed to the imperfect functional form used in modelling the observed cluster mass profile with the present of, e.g., halo triaxiality or substructures.
The resulting weak-lensing mass bias is obtained as $\bwl = 0.99\pm0.05$ for the shear-selected cluster sample, on average (see Section~\ref{sec:modelling_bwl_bias}).
This constraint is used as the informative prior on \bwl\ in the forward modelling.

The measurement uncertainty of the weak-lensing observable \snr, together with the selection function, is quantified and accounted for by using an injection-based method, of which the details are described in Section~\ref{sec:modelling_selection_function} and fully given in the companion paper \citep{chen24}.
In short, we parameterize the distribution of \snr\ at a given weak-lensing mass \mwl\ by the aperture mass peak height \mkappa\ and the angular size \thetas\ of the scale radius.
The former (\mkappa) and latter (\thetas) describe the normalization and the extendedness of a mass profile, respectively.
By injecting synthetic clusters into the observed aperture-mass maps, we determine the distribution $P\left(\snr | \mkappa, \thetas\right)$ and the sample completeness $\comp\left(\mkappa,\thetas \right)$ of the shear-selected clusters given the selection $\nu\geq4.7$.
This injection-based method naturally accounts for all the observational uncertainties, including the miscentering, the shape noise, the variation of the imaging depth, the masking due to the survey boundary or bright stars, and most importantly, the cosmic-shear noise originated from large-scale structures.

To better quantify the measurement uncertainty in \snr\ with respect to the intrinsic one in the injections of synthetic clusters, we introduce a selection parameter \deltasel\ in quantifying $\comp\left(\mkappa,\thetas \right)$ and $P\left(\snr | \mkappa, \thetas\right)$.
We marginalize the parameter \deltasel\ in the cosmological constraints over a range that is informed by the optical counterparts of the shear-selected clusters (see Section~\ref{sec:injection}).

We quantify the photo-\redshift\ bias of the sources by either the technique of halo clustering (Section~\ref{sec:clustering_deltaz}) or that informed by the HSC-Y3 cosmic-shear analyses (Section~\ref{sec:cosmic_shear_deltaz}).
The former implies no photo-\redshift\ bias, while the latter suggests the presence of the photo-\redshift\ bias at a level of $\approx2\sigma$.
We take into account the photo-\redshift\ bias and find no impact on our cosmological constraints, if it exists.

In a blinded analysis, we obtain the fully marginalized posteriors of
cosmological parameters in a flat \lcdm\ model as
$\omegam = \omegamDefault$,
$\sigmaeight = \sigmaeightDefault$,
$\seightnohat \equiv \sigmaeight\sqrt{ \omegam/0.3 } = \seightNormDefault$, and 
\[
\seight \equiv\sigmaeight\left(\omegam/0.3\right)^{0.25} = \seightDefault \, .
\]
In this work, we can only put a statistically meaningful constraint on \seight, while the others suffer from severe parameter degeneracy.
We extensively examine and find our results robust against several systematics, including analysis choices made in accounting for the weak-lensing mass bias and photo-\redshift\ bias, the fitting formulae of the halo mass function, the selection threshold \numin, and the difference among individual subfields.

We compare our cosmological constraints with other studies, including those based on the abundance of clusters selected using a baryonic tracer (in the optical, X-rays, and the mm wavelength), the HSC-Y3 cosmic shears, the $3\times2$pt analyses (from HSC-Y3, DES, and KiDS), and the \planck\ CMB observations.
Our result is in excellent agreement with them at a level of $\lesssim1\sigma$, except for the mild but not statistically significant discrepancy with the $3\times2$pt analyses at levels of $\approx2\sigma$. 
No \seightnohat\ tension is seen between the abundance of the shear-selected clusters and the \planck\ CMB observation.
Interestingly, we find that our result shows excellent agreement with the HSC-Y3 cosmic-shear analyses, which utilizes the same data sets as in this work, but reveals the discrepancy with the HSC-Y3 $3\times2$pt analysis at a level of $\approx2\sigma$.
This suggests that the analysis of galaxy-galaxy clustering might play a key role in resolving the \seightnohat\ tension between \planck\ and the $3\times2$pt analyses.

To sum up, we present the cosmological constraints from cluster abundance leveraging the largest weak-lensing shear-selected sample to date, with a novel strategy developed in the companion paper \cite{chen24} to model the selection function directly on the observed aperture-mass maps.
This work promises the success of utilizing shear-selected clusters as a cosmological probe with high-quality data in the imminent era of wide-field weak-lensing surveys, e.g., the Legacy Survey of Space and Time (LSST) operated by Rubin Observatory \citep{ivezic19}, the Nancy Grace Roman Space Telescope \citep{spergel15}, and the \textit{Euclid} mission \citep{laureijs11,euclid24}.

%
%

\section*{Acknowledgements}

We thank the anonymous referee for constructive comments that lead to improvements in this paper.
I-Non Chiu acknowledges the support from the National Science and Technology Council in Taiwan (Grant NSTC 111-2112-M-006-037-MY3).
This work made use of the computational and storage resources in the Academic Sinica Institute of Astronomy and Astrophysics (ASIAA) and the National Center for High-Performance Computing (NCHC) in Taiwan.
I-Non Chiu thanks the hospitality of Ting-Wen Lan and Ji-Jia Tang at National Taiwan University (NTU) and Alex Saro at the University of Trieste.
Kai-Feng Chen acknowledges support from the Taiwan Think Global Education Trust Scholarship and the Taiwan Ministry of Education's Government Scholarship to Study Abroad.
Kai-Feng Chen thanks the hospitality of Yen-Ting Lin at ASIAA and Adrian Liu at McGill University.
Yen-Ting Lin acknowledges supports by the grant NSTC 112-211-M-001-061.
This work was supported by JSPS KAKENHI Grant Numbers JP20H05856, JP22H01260, JP22K21349.

The Hyper Suprime-Cam (HSC) collaboration includes the astronomical communities of Japan and Taiwan, and Princeton University. The HSC instrumentation and software were developed by the National Astronomical Observatory of Japan (NAOJ), the Kavli Institute for the Physics and Mathematics of the Universe (Kavli IPMU), the University of Tokyo, the High Energy Accelerator Research Organization (KEK), the Academia Sinica Institute for Astronomy and Astrophysics in Taiwan (ASIAA), and Princeton University. Funding was contributed by the FIRST program from the Japanese Cabinet Office, the Ministry of Education, Culture, Sports, Science and Technology (MEXT), the Japan Society for the Promotion of Science (JSPS), Japan Science and Technology Agency (JST), the Toray Science Foundation, NAOJ, Kavli IPMU, KEK, ASIAA, and Princeton University. 

This paper makes use of software developed for Vera C. Rubin Observatory. We thank the Rubin Observatory for making their code available as free software at \url{http://pipelines.lsst.io/}.

This paper is based on data collected at the Subaru Telescope and retrieved from the HSC data archive system, which is operated by the Subaru Telescope and Astronomy Data Center (ADC) at NAOJ. Data analysis was in part carried out with the cooperation of Center for Computational Astrophysics (CfCA), NAOJ. We are honored and grateful for the opportunity of observing the Universe from Maunakea, which has the cultural, historical and natural significance in Hawaii. 

The Pan-STARRS1 Surveys (PS1) and the PS1 public science archive have been made possible through contributions by the Institute for Astronomy, the University of Hawaii, the Pan-STARRS Project Office, the Max Planck Society and its participating institutes, the Max Planck Institute for Astronomy, Heidelberg, and the Max Planck Institute for Extraterrestrial Physics, Garching, The Johns Hopkins University, Durham University, the University of Edinburgh, the Queen’s University Belfast, the Harvard-Smithsonian Center for Astrophysics, the Las Cumbres Observatory Global Telescope Network Incorporated, the National Central University of Taiwan, the Space Telescope Science Institute, the National Aeronautics and Space Administration under grant No. NNX08AR22G issued through the Planetary Science Division of the NASA Science Mission Directorate, the National Science Foundation grant No. AST-1238877, the University of Maryland, Eotvos Lorand University (ELTE), the Los Alamos National Laboratory, and the Gordon and Betty Moore Foundation.

This work is possible because of the efforts in the Vera C. Rubin Observatory \citep{juric17,ivezic19} and PS1 \citep{chambers16, schlafly12, tonry12, magnier13}, and in the HSC \citep{aihara18a} developments including the deep imaging of the COSMOS field \citep{tanaka17}, the on-site quality-assurance system \citep{furusawa18}, the Hyper Suprime-Cam \citep{miyazaki15, miyazaki18, komiyama18}, the design of the filters \citep{kawanomoto18},  the data pipeline \citep{bosch18}, the design of bright-star masks \citep{coupon18}, the characterization of the photometry by the code \texttt{Synpipe} \citep{huang18}, the photometric redshift estimation \citep{tanaka18}, the shear calibration \citep{mandelbaum18}, and the public data releases \citep{aihara18b, aihara19}.

This work made use of the IPython package \citep{ipython}, \texttt{SciPy} \citep{scipy}, \texttt{TOPCAT}, an interactive graphical viewer and editor for tabular data \citep{topcat1,topcat2}, \texttt{matplotlib}, a Python library for publication quality graphics \citep{matplotlib}, \texttt{Astropy}, a community-developed core Python package for Astronomy \citep{astropy}, \texttt{NumPy} \citep{van2011numpy}, and \texttt{Pathos} \citep{pathos} a utility of parallel computing. 
The corner plots for the parameter constraints are produced by \texttt{ChainConsumer} \citep{hinton2016}.
The code \texttt{pyccl} \citep{chisari19} is used to calculate cosmology-related quantities. 
The software management used in this work leverages the \texttt{conda-forge} project \citep{conda_forge_community_2015_4774216}.

%
%

\section*{Data Availability}

The weak-lensing shape catalogs used in this work will be released in the Public Data Release of the HSC-Y3 data sets.
The chains of the cosmological parameters are publicly accessible via \url{https://github.com/inonchiu/hsc_shear_selected_clusters}. 
The other data products underlying this article will be shared upon a reasonable request to the corresponding author.

%
%

\bibliographystyle{aasjournal}
\bibliography{literature} 

%
%

\onecolumngrid

\appendix

\section{The semi-analytical injection simulations}
\label{app:injection}

In what follows, we describe the methodology of the injection-based approach to quantify the selection function, including the sample completeness $\comp\left(\mkappa,\thetas \right)$ and the distribution $P_{\numin}\left(\snr | \mkappa,\thetas  \right)$.
Again, we refer readers to the companion paper \citep{chen24} for more details.

First, we generate a mock cluster catalog in a flat \lcdm\ cosmology with the cosmological parameters of $\left(\Hnow, \omegam\right) = \left(70~\frac{k\mathrm{m}/s}{\Mpc}, 0.3\right)$.
In the interest of covering a wide range of \mkappa\ and \thetas\ with enough statistics, we generate $\approx2.5$ millions synthetic clusters by uniformly sampling the halo mass \Mtwooo, the redshift \zcl, and the angular size \thetas\ in the ranges of $5\times10^{12}<\frac{ \Mtwooo }{ \Msunh }<5\times10^{16}$, $0.01 < \zcl < 2$, and $0.01 < \frac{\thetas}{\mathrm{arcmin}} < 30$~arcmin, respectively.
Given the sampled $\left(\Mtwooo, \zcl, \thetas\right)$, the projected mass profile of each cluster is evaluated by assuming a spherical NFW model (see Section~\ref{sec:wlbasics}).
We aim to determine the probability $P_{\numin}\left(\snr | \mkappa,\thetas  \right)$ and the completeness $\comp\left(\mkappa,\thetas \right)$ over a wide range of \mkappa\ and \thetas\ by directly sampling $\left(\Mtwooo, \zcl, \thetas\right)$, therefore the sampling between the weak-lensing mass \mwl\ and the halo mass \Mtwooo\ is not required.

Next, we inject the lensing signal of the synthetic clusters into the weak-lensing aperture-mass maps.
The injection is performed at the catalog level.
For a synthetic cluster, we randomly assign a sky position as the cluster center.
For a source galaxy within the clustercentric radius of $\rdd < 80~\Mpch$, the resulting distortion in the ellipticity of the source is 
\begin{equation}
\label{eq:cluster_injection_e}
e_{\mathrm{clu}} = 2\mathcal{R}\left(1 + m\right) g_{\mathrm{clu}} \, ,
\end{equation}
where $\mathcal{R}$ and $m$ are the response and the multiplicative bias of the source galaxy in the HSC shape catalog, respectively, and $g_{\mathrm{clu}}$ is the tangential reduced shear  
induced by the synthesis cluster at the sky position.
When calculating the reduced shear, we make use of the photo-\redshift\ estimation of the source.
Specifically, for each source we use the redshift point estimate $\redshift_{\mathrm{MC}}$, which is randomly sampled from the full redshift distribution of the source, to calculate the critical surface mass density (equation~(\ref{eq:sigmacrit})).
We assume a weak-lensing regime, so that the ellipticities before ($e_{\mathrm{before}}$) and after ($e_{\mathrm{after}}$) the injection are related to each other as
\begin{equation}
\label{eq:injection_ellip}
e_{\mathrm{after}} = e_{\mathrm{before}} + e_{\mathrm{clu}} \, .
\end{equation}
In this way, the post-injection shear $g_{\mathrm{after}}$ reads
\begin{equation}
\label{eq:injection_g}
g_{\mathrm{after}}
= \frac{1}{\left(1 + m\right)}\left(\frac{ e_{\mathrm{after}} }{2\mathcal{R}} - c\right)
= g_{\mathrm{before}} + g_{\mathrm{clu}} \, ,
\end{equation}
where $c$ is the additive bias, and $g_{\mathrm{before}}$ is the lensing signal of cosmic structures before the injection (i.e., the cosmic shears) evaluated as
$g_{\mathrm{before}} = \frac{1}{\left(1 + m\right)}\left(\frac{ e_{\mathrm{before}} }{2\mathcal{R}} - c\right)$.

Note that we do not use the full equation of the lensing distortion in injecting the synthetic signal of shears \citep[i.e., equation~(24) in][see also \citealt{bernstein02}]{shirasaki19}.
This is because the formulae of the lensing distortion in \cite{shirasaki19} describe the transform of the ellipticity from the intrinsic one without any pre-existing lensing signals, while we inject the synthetic signals into the maps where the observed signals of the cosmic shears are already present and intentionally retained.

Finally, after the injection we produce the aperture-mass maps following the same procedure as described in Section~\ref{sec:practical}, perform the peak finding of the injected synthesis clusters, and quantify the completeness $\comp\left(\mkappa,\thetas\right)$ and the distribution $P_{\numin}\left(\snr |\mkappa,\thetas\right)$ as functions of the intrinsic aperture mass peak \mkappa\ and the angular size \thetas.
To derive $\comp\left(\mkappa,\thetas\right)$ and $P_{\numin}\left(\snr |\mkappa,\thetas\right)$,
we bin the injected clusters into fine grids in the space of \mkappa\ and \thetas.
Because the injected clusters are sampled uniformly in the space of $\left(\mwl, \zcl, \thetas\right)$, this results in uneven numbers in the \mkappa-\thetas\ grids.
In the range of interested, $1\lesssim\mkappa\lesssim30$ and $0.03\lesssim\frac{\thetas}{\mathrm{arcmin}}\lesssim30$, the number of injected clusters in a \mkappa-\thetas\ grid ranges from $\approx100$ to $\approx800$ with an average of $\approx600$. 
This gives satisfying statistical power in deriving $\comp\left(\mkappa,\thetas\right)$ and $P_{\numin}\left(\snr |\mkappa,\thetas\right)$.

\section{The derivation of the constant weak-lensing mass bias}
\label{app:derive_bwl}

In the following, we describe the procedure to infer the constant weak-lensing mass bias of the shear-selected clusters based on the optical richness of their optical counterparts.

Firstly, we identify the optical counterparts of the shear-selected clusters by cross-matching them with those from the \texttt{CAMIRA} optical cluster catalog \citep{oguri14,oguri18}, which is constructed using the red-sequence-based cluster finding algorithm \citep{gladders00} on the HSC S21A data.
This gives the redshift \redshift\ and the optical richness \rich\ of the individual shear-selected clusters.
Note that the photometric redshift of \texttt{CAMIRA} clusters was closely examined in \cite{tcchen24}, demonstrating a high accuracy to construct samples of superclusters. 
The methodology of the cross-matching is described in detail in the companion paper \citep{chen24}.
In short, we assess the cross-matching by the probability $P\left( \rich, \redshift | \snr \right)$ of observing the richness \rich\ of such an optical counterpart (within a positional offset of $<8~\mathrm{arcmin}$) at the redshift \redshift\ given the lensing peak height \snr\ of the shear-selected cluster.
We include the intrinsic scatter of the richness at a fixed halo mass in calculating the probability $P\left( \rich, \redshift | \snr \right)$.
Out of $129$ shear-selected clusters with $\snr\geq4.7$, there are $106$ systems (corresponding to a matching rate of $\approx82\percent$) with matched optical counterpart in the \texttt{CAMIRA} catalog at $\redshift>0.1$,
Including other low-\redshift\ catalogs of optically selected clusters in the cross-matching leads to a sample purity of $\gtrsim99\percent$.
That is, the probability of a shear-selected cluster being a line-of-sight alignment by chance is less than $1\percent$.

Secondly, we estimate the mass \mass\ of the shear-selected clusters based on the richness \rich\ and redshift \redshift\ of the optical counterparts by inverting the richness-to-mass-and-redshift (\rich--\mass--\redshift) relation.
We only consider those shear-selected clusters with \texttt{CAMIRA} optical counterparts.
There are three independent studies on constraining the \rich--\mass--\redshift\ relation of \texttt{CAMIRA} clusters---\cite{murata19} based on a joint analysis of the cluster abundance and weak shear, \cite{chiu20} using the effect of weak-lensing magnification, and \cite{chiu20b} leveraging the properties of halo clustering.
In each work, they assumed that the richness \rich\ follows a log-normal distribution with the intrinsic scatter \sigmarich\ around a mean richness predicted by the \rich--\mass--\redshift\ relation, which has a mass scaling with the power-law index \Brich.
To estimate the mass of a shear-selected cluster based on the optical counterpart, we sample the cluster mass $\mass_{\mathrm{sampled}}$ following a log-normal distribution with the intrinsic scatter $\frac{\sigmarich}{\Brich}$ around a mean mass, which is obtained by inverting the \rich--\mass--\redshift\ relation given the richness \rich\ and redshift \redshift\ of the counterpart.
For each shear-selected cluster, we then calculate the corresponding weak-lensing mass bias 
${\bwl}_{,\mathrm{sampled}} \equiv \exp{ \left\langle \ln\left(\bwl | \mass_{\mathrm{sampled}},\redshift \right)\right\rangle}$ 
given the sampled mass $\mass_{\mathrm{sampled}}$ at the cluster redshift \redshift\ using 
equation~(\ref{eq:wlmass2halomass}).
The constant weak-lensing mass bias of the shear-selected sample is determined as the mean value of $\left\langle{\bwl}_{,\mathrm{sampled}}\right\rangle$ among all clusters in the sample.
We perform the same procedure separately for the \rich--\mass--\redshift\ relations obtained in \cite{murata19}, \cite{chiu20}, and \cite{chiu20b}.

Finally, the process mentioned above is repeated for $1000$ random realizations of 
the \rich--\mass--\redshift\ relation parameters and 
the \mwl--\mass--\redshift\ relation parameters $\left( \Awl, \Bwl, \gammawl\right)$.
This results in a distribution of the constant weak-lensing mass bias from the $1000$ realizations.
In a Gaussian approximation, the mean of the distribution is,
$0.996 \pm 0.043$,
$0.999 \pm 0.044$, and
$0.972 \pm 0.048$
for the \rich--\mass--\redshift\ relation obtained from \cite{murata19}, \cite{chiu20}, and \cite{chiu20b}, respectively.
We derive the mean of these results based on the three \rich--\mass--\redshift\ relations, leading to equation~(\ref{eq:wlpriors_constant}) used as the informative prior on \Awl.

\section{The constraints of all parameters}
\label{app:auxiliary}

\begin{figure*}
\centering
\resizebox{\textwidth}{!}{
\includegraphics[scale=1]{
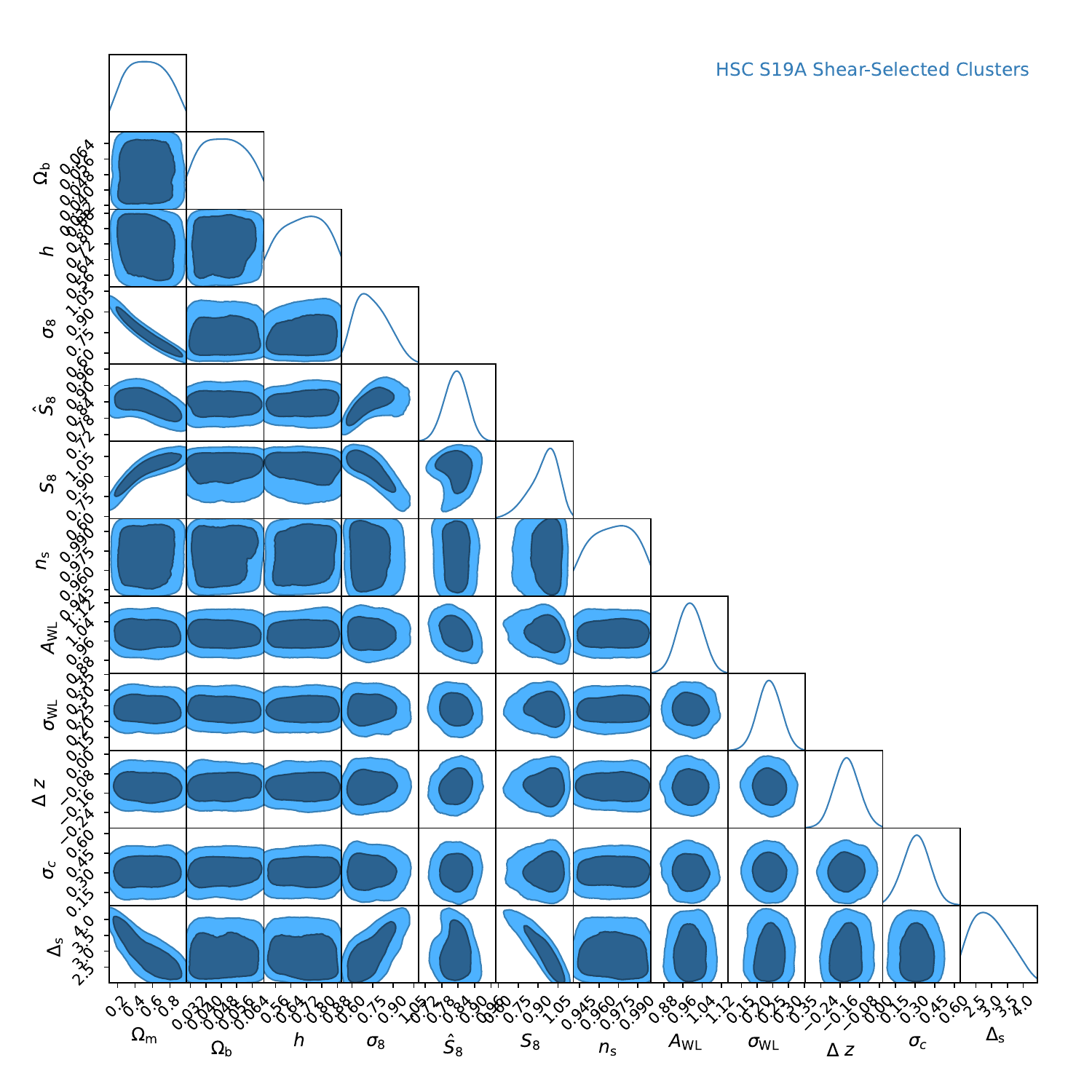
}}
\caption{
The posteriors of and covariances between all free parameters in the default analysis. 
The same plotting configuration used in generating Figure~\ref{fig:gtc_som} is also adopted here.
}
\label{fig:gtc_all}
\end{figure*}

\begin{figure*}
\centering
\resizebox{0.30\textwidth}{!}{
\includegraphics[scale=1]{
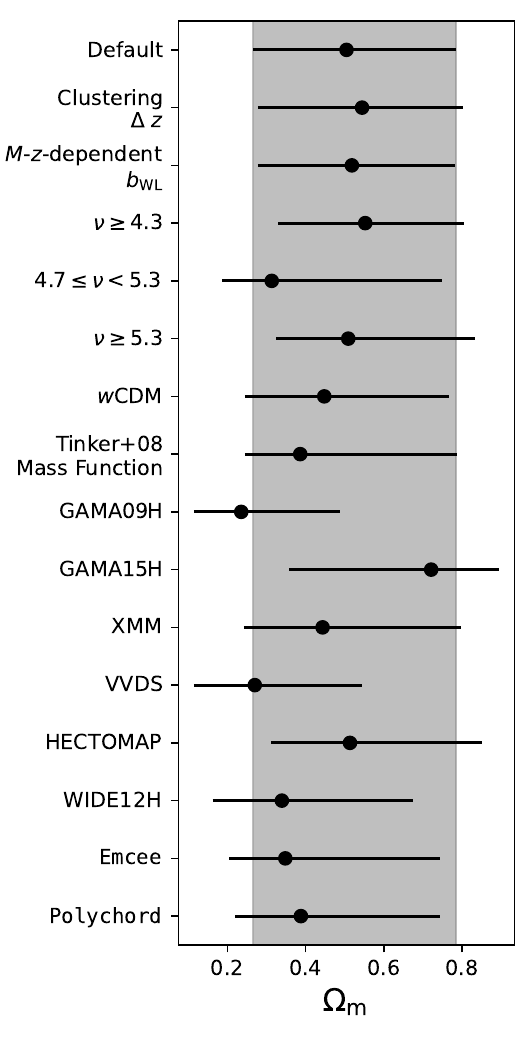
}}
\resizebox{0.30\textwidth}{!}{
\includegraphics[scale=1]{
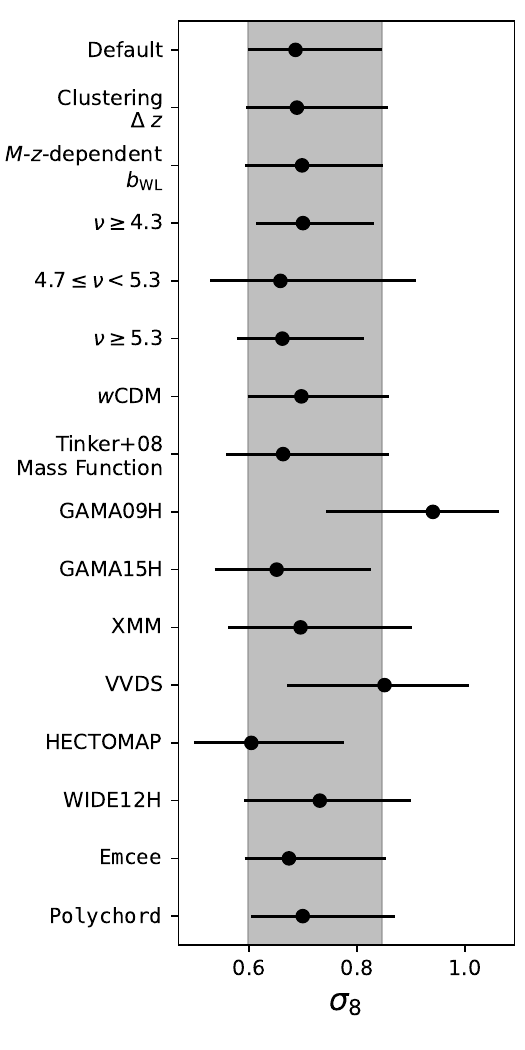
}}
\resizebox{0.30\textwidth}{!}{
\includegraphics[scale=1]{
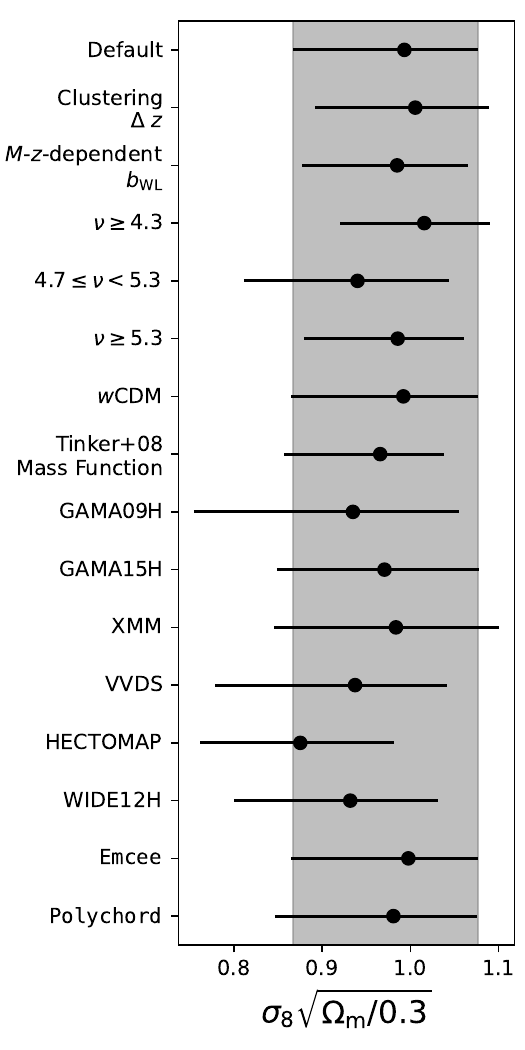
}}
\caption{
The systematic uncertainty in the marginalized posterior of \omegam\ (\textit{left}), \sigmaeight\ (\textit{middle}) and, \seightnohat\ (\textit{right}).
These plots are made using the same configuration as in Figure~\ref{fig:forest}.
}
\label{fig:forest2}
\end{figure*}

In Figure~\ref{fig:gtc_all}, we show the fully marginalized posteriors and covariance of all the free parameters used in the default analysis. 
The fully marginalized posteriors of the cosmological parameters (\omegam, \sigmaeight, and \seightnohat) against various systematic uncertainties are plotted in Figure~\ref{fig:forest2}.

%


\label{lastpage}
\end{document}